%% file: neural_sensorimotor_arxiv_v2.tex
\documentclass{article}

\input{_macros}
\input{_definitions}

\begin{document}
	
\title{ \bf Neural Models and Algorithms for Sensorimotor Control \\ of an Octopus Arm
	\vspace{10pt}} 
\author{
	Tixian Wang$^{1,2}$, Udit Halder$^{2,}$\footnote{Corresponding author, email: {\color{blue} \texttt{udit@illinois.edu}}}\,\,, Ekaterina~Gribkova$^{3}$, \\Rhanor Gillette$^{3,4}$, Mattia Gazzola$^{1,5,6}$, and Prashant G.~Mehta$^{1,2}$
	\vspace{10pt}
}
\date{
	\normalsize
	$^{1}$Department of Mechanical Science and Engineering,\\
	$^{2}$Coordinated Science Laboratory, \\
	$^{3}$Neuroscience Program, \\
	$^{4}$Department of Molecular and Integrative Physiology,\\
	$^{5}$National Center for Supercomputing Applications, \\
	$^{6}$Carl R. Woese Institute for Genomic Biology, \\
	University of Illinois Urbana-Champaign, IL, 61801, USA
}

\maketitle
\thispagestyle{empty}

\vspace{40pt}
\begin{abstract}
	In this article, a biophysically realistic model of a soft octopus arm with internal musculature is presented. The modeling is motivated by experimental observznations of sensorimotor control where an arm localizes and reaches a target. Major contributions of this article are: (i) development of models to capture the mechanical properties of arm musculature, the electrical properties of the arm peripheral nervous system (PNS), and the coupling of PNS with muscular contractions; (ii) modeling the arm sensory system, including chemosensing and proprioception; and (iii) algorithms for sensorimotor control, which include a novel feedback neural motor control law for mimicking target-oriented arm reaching motions, and a novel consensus algorithm for solving sensing problems such as locating a food source from local chemical sensory information (exogenous) and arm deformation information (endogenous). Several analytical results, including rest-state characterization and stability properties of the proposed sensing and motor control algorithms, are provided. Numerical simulations demonstrate the efficacy of our approach. Qualitative comparisons against observed arm rest shapes and target-oriented reaching motions are also reported.
	
	\medskip
	\noindent
	\textit{Subject Areas} -- applied mathematics, biomechanics, mathematical modeling, nonlinear systems 
	
\end{abstract}

\clearpage
\pagenumbering{arabic} 

%
%

\section{Introduction}\label{sec:introduction}

This paper is concerned with developing a bioinspired model of an octopus arm. There are two distinguishing aspects: (1) sub-models are described for the flexible arm, including the actuation (muscles), the nervous system, and sensing modalities, and (2) a sensorimotor control law is described for the arm to sense and reach a target in the environment. Our work represents the first such neurophysiologically plausible end-to-end (from sensory input to arm motion) model of an octopus arm. For the sake of clarity and exposition, the model is described here for a planar (2D) setting.

\begin{figure*}[t]
	\centering
	\includegraphics[width=\textwidth, trim = {0pt 0pt 0pt 0pt}, clip = false]{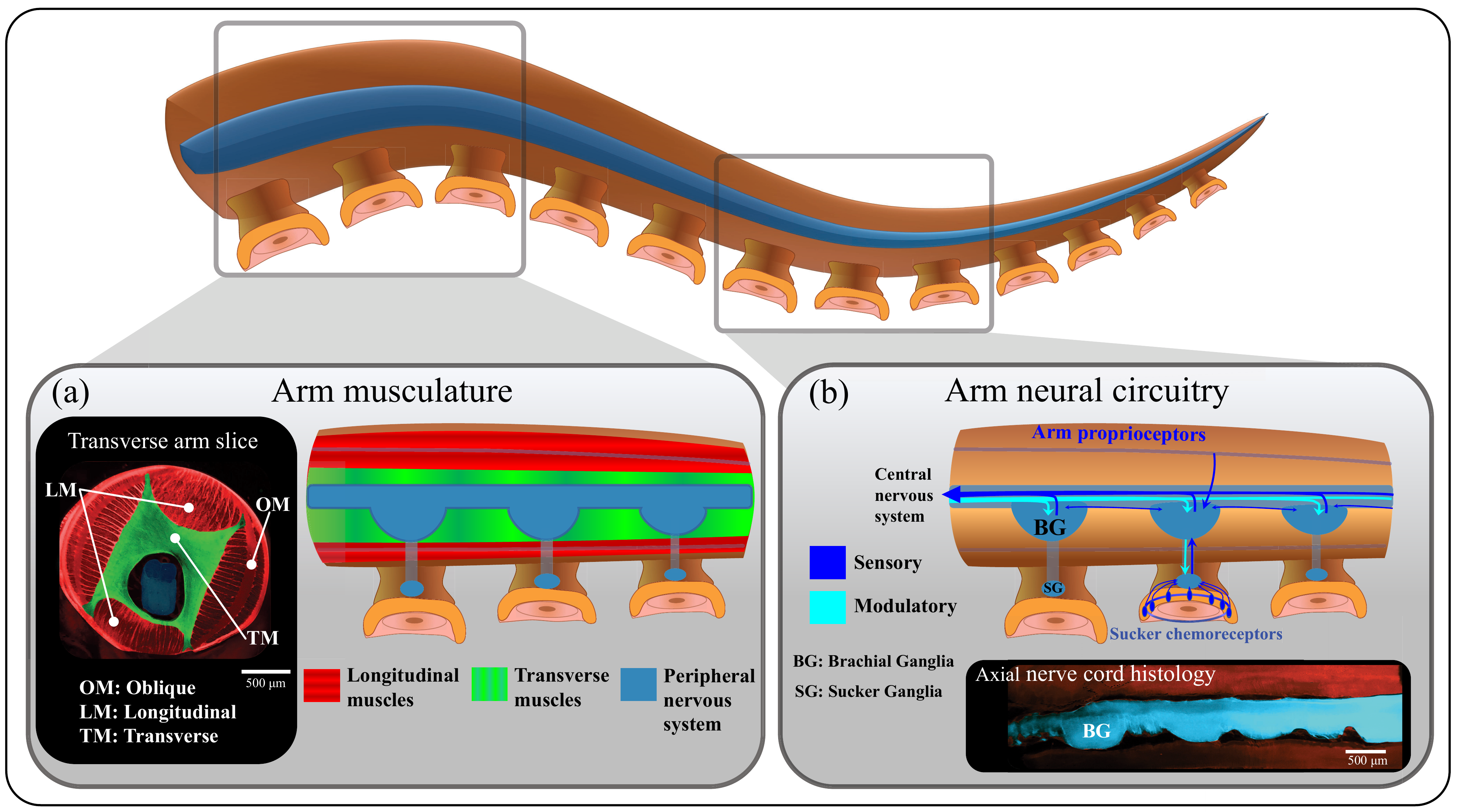}
	\vspace{-20pt}
	\caption{Biophysiology of an octopus arm: (a) Cross-sectional views of the arm musculature showing the peripheral nervous system (in blue), and the prominent muscle groups including longitudinal (in red) and transverse muscles (in green); (b) Anatomy of the sensorimotor control system following\cite{grasso2014octopus} with histology of the axial nerve cord in the inset.}
	\label{fig:biophysiology}
	\vspace{-10pt}
\end{figure*}

\paragraph{Biophysiology of an octopus arm.}
The left panel of Figure~\ref{fig:biophysiology}(a) depicts the cross-section of an isolated arm. Its innermost structure is an axial nerve cord (ANC) which runs along the centerline of the arm. The ANC is surrounded by densely packed muscles which are of the following three types~\cite{kier2007arrangement, kier2016musculature}: (a) transverse muscles (TMs) are arranged radially along the cross-section; (b) longitudinal muscles (LMs) run parallel to the centerline; and (c) oblique muscles (OMs) are arranged in a helical fashion. Along the length of the arm are a discrete arrangement of suckers (see Figure~\ref{fig:biophysiology}(b)). Each sucker has an abundance of sensory receptors ($\sim10^4$), including chemosensory and mechanosensory cells~\cite{graziadei1976sensory,mather2021octopus}. Stretch receptors are located inside the intramuscular nerve cords and are believed to sense local arm deformations~\cite{grasso2014octopus, matzner2000neuromuscular}. Sensing and actuation are coordinated through the arm's peripheral nervous system (PNS) which includes the ANC. Located beneath each sucker is a {\em sucker ganglion} that receives sensory input from the sensors and sends motor commands to the sucker muscles, controlling the orientation of the sucker~\cite{grasso2014octopus}. Associated with each sucker ganglion is a {\em brachial ganglion} which integrates sensory information~\cite{grasso2014octopus} and sends motor commands along numerous nerve roots to actuate the arm muscles~\cite{matzner2000neuromuscular}. These ganglia are thought to act as `\textit{mini brains}' for the arm where much of the local sensorimotor computations take place~\cite{grasso2014octopus}.

\begin{figure*}[t]
	\centering
	\includegraphics[width=\textwidth, trim = {0pt 0pt 0pt 0pt}, clip = false]{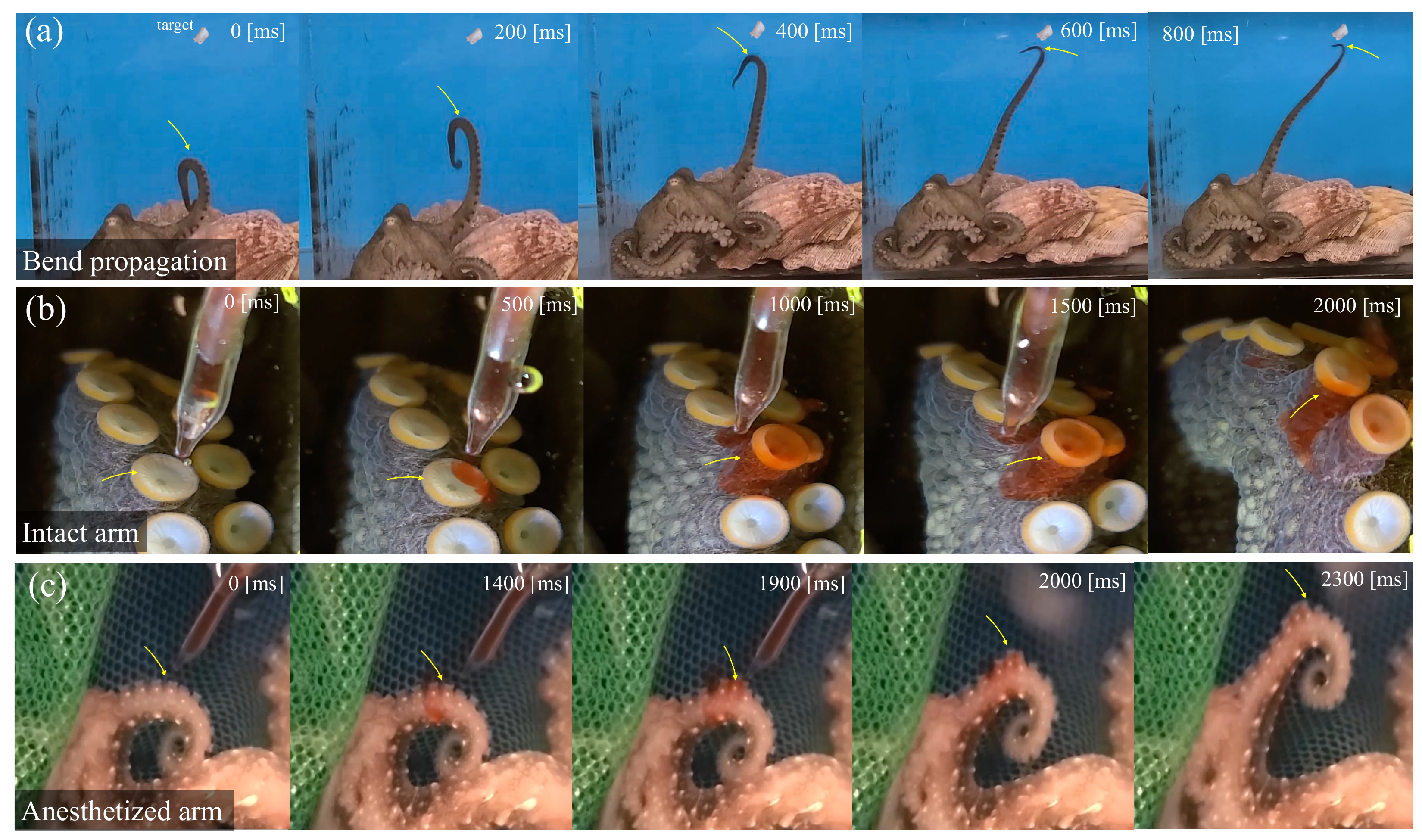}
	\vspace{-20pt}
	\caption{Behavioral experiments on \textit{Octopus bimaculoides} and \textit{Octopus rubescens}: (a) A video sequence of an arm performing bend propagation. Yellow arrows indicate the bend location; (b) Timelapse of an intact arm suckers reacting to chemical stimulus with yellow arrows indicating the reorientation of the sucker; (c) Timelapse of an anesthetized arm reacting to chemical stimulus. Yellow arrows show the part of the arm that is most reactive to the stimulus and forms a localized bend.}
	\label{fig:behavior}
	\vspace{-10pt}
\end{figure*}

\paragraph{Behavioral observations that motivate the modeling.} The prowess of an octopus arm to execute sensorimotor control strategies has few parallels in the natural world. An arm has virtually infinite degrees of freedom which an octopus is able to manipulate through a distributed control of the internal muscles. An illustration of the same appears in Figure~\ref{fig:behavior}(a) which depicts the successive frames of a freely moving octopus arm as it reaches a target (shrimp). This type of motion primitive is referred to as {\em bend propagation} wherein a bend is created near the base and \textit{actively} propagated along the length of the arm through a traveling wave of muscle actuation~\cite{gutfreund1998patterns, sumbre2001control, hanassy2015stereotypical}. The goal-directed motion of the arm is informed by distributed sensing from the chemosensors and proprioceptors. In this paper, we make a distinction between two types of sensing: (a) sensing of external variables such as the location of the target and (b) sensing of internal variables such as the shape of the arm. Experimental evidence for external sensing is depicted in Figure~\ref{fig:behavior}(b)-(c) showing the arm's ability to sense the direction (bearing) of the chemical stimulus. Additional details on both these experiments appear as part of~\ref{appdx:experiments}.

\paragraph{Contributions of this work.} Inspired by behavioral experiments, a mathematical sensorimotor control problem is defined in \S\,\ref{sec:math_problem}. The remainder of the paper presents a solution to this problem. There are three aspects to our solution:

\vspace*{-5pt}
\begin{enumerate}
	\item
	\textit{Modeling the neuromusculature and control algorithm.} 
	A contribution of our work is the control-oriented biophysical modeling of the known neuromusculature: (a) Cosserat rod theory~\cite{antman1995nonlinear} is used to model the flexible aspects (nonlinear elasticity) of the arm dynamics; (b) a continuum extension of the Hill's theory of muscles~\cite{hill1938heat} is used to model the distributed forces and couples produced by the arm's three main muscle groups; 
	and (c) cable theory~\cite{tuckwell1988introduction} is employed to model the PNS that provides the neural activation of the muscles. Another important contribution is a novel feedback control law that provides the input to the PNS. Using this control law, the feedback control system is shown to solve the problem of reaching a stationary target (Theorem~\ref{thm:sensoryfeedback_control}).
	
	\item
	\textit{Modeling the sensory system and estimation algorithm.}
	To capture the neuroanatomy of the arm, a distributed but \textit{discrete} architecture for sensing is proposed, where the neural computing machinery is located at each node (ganglia). These independent computing units {communicate} with each other in order to solve two kinds of sensing problems: locating the target based on measuring only local chemical concentrations (chemosensing) and estimating the arm shape based on arm curvature measurement (proprioception). Models for the neural circuitries for sensing are described using the theory of neural rings~\cite{zhang1996representation, hahnloser2003emergence}. 
	The inputs to the neural rings are informed by a novel consensus-type estimation algorithm whose convergence analysis is provided and shown to solve the sensing problems (Theorem~\ref{thm:consensus}).
	
	\item
	\textit{Numerical simulations.} 
	The mathematical models are presented in a modular fashion which allows us to validate and illustrate each piece of the model with analytical results and corresponding numerical simulations. 
	The simulations are carried out using the software environment \textit{Elastica}~\cite{gazzola2018forward, zhang2019modeling}. Detailed description of the simulation environment setup appears in~\ref{appdx:numerics_setup} and is summarized in Table~\ref{tab:numerics}. 
\end{enumerate}

The models of the neuromusculature aim at capturing salient features of known biophysics of an octopus arm. The models for sensing and control algorithms are bio-inspired but no claims are made concerning their pertinence to the neural mechanisms of an octopus arm. Indeed, much of the details of the octopus arm neural circuitry remain undiscovered. 

\medskip 
\noindent 
{\it Bioinspiration of sensing and control.}
An important feature of this paper is that both the feedback control law and the neural rings are inspired from biology. In particular, the proposed control law takes inspiration from biologically plausible control strategies in nature, e.g. motion camouflage (in bats, falcons)~\cite{glendinning2004mathematics, ghose2006echolocating, justh2006steering} or classical pursuit (in flies, honey bees)~\cite{wei2009pursuit, galloway2013symmetry, halder2016steering}. Next, the use of neural rings for sensing is motivated by the theory of head direction cells in animals (e.g., mice)~\cite{zhang1996representation, hahnloser2003emergence}. These connections are explained as Remark~\ref{remark:motion_camouflage} and Remark~\ref{remark:neural_ring} in the paper. 

Our work, based on rigorous definition of the mathematical problem and analysis of the proposed solution, is expected to provide an impetus to both biologists and soft roboticists. For example, proposed models for sensing based on neural ring architecture can be tested with experiments, offering the possibility of new understanding of the underlying working mechanisms of octopus arms. On the other hand, both modeling and sensorimotor control algorithms are expected to be useful for design and control of octopus-inspired soft robotics which has been a focus of recent studies~\cite{guglielmino2010octopus, renda2014dynamic, xie2023octopus}.

\paragraph{Literature survey.}
Biophysical modeling of animal locomotion has a rich history~\cite{dickinson2000animals, ijspeert2008central, rossignol2006dynamic,holmes2006dynamics, ramdya2023neuromechanics, matsuoka1984dynamic}, ranging from lampreys \cite{ekeberg1993combined}, salamanders~\cite{ijspeert2007swimming}, zebrafishes \cite{chemtob2020strategies}, and fruit flies~\cite{namiki2018functional}, to humans~\cite{geyer2010muscle}. A recent focus has been on translating these efforts into models of robots, actuators, and control~\cite{ijspeert2007swimming, aydin2019neuromuscular, folgheraiter2019neuromorphic, polykretis2023bioinspired}.  
Seminal studies and experimental observations from biologists provide an important motivation for our modeling work~\cite{gutfreund1998patterns, sumbre2001control, yekutieli2005dynamic, hanassy2015stereotypical, flash2023biomechanics}. An early work on modeling of an octopus arm is based on discrete mass-spring systems~\cite{yekutieli2005dynamic, yekutieli2005dynamic2}.
In our own prior work, we have used Cosserat rod theory to model and simulate the arm dynamics including its musculature~\cite{chang2021controlling, chang2023energy, tekinalp2023topology}. Several types of control strategies have been investigated for octopus arm movements, including stiffening wave actuation~\cite{yoram2002move, yekutieli2005dynamic, yekutieli2005dynamic2, wang2022control}, energy shaping control~\cite{chang2020energy, chang2021controlling, chang2023energy}, optimal control~\cite{cacace2019control,cacace2020modeling, wang2021optimal}, hierarchical control~\cite{shih2023hierarchical}, and feedback control~\cite{wang2022sensory, wang2022modeling}. Apart from control, sensing in octopus arms has also been studied by biologists~\cite{hanlon2018cephalopod, van2020molecular, kang2023sensory, polese2016olfactory}. However, mathematical models of sensing remain scarce in the literature~\cite{ishida2021model}. In this paper, we provide the first such sensing models and moreover integrate these with models of the PNS as well as novel algorithms of estimation and control.

\paragraph{Outline.}
The remainder of the paper is organized as follows. A mathematical formulation of the sensorimotor control problem is given in \S\,\ref{sec:math_problem}.   
The control and sensing architectures are presented in \S\,\ref{sec:control_models} and \S\,\ref{sec:sensing_models}, respectively. Each of these sections also include algorithms and analytical results for the same. The end-to-end sensorimotor control algorithm is obtained by combining the two in \S\,\ref{sec:algorithm_combined}.
The paper concludes in \S\,\ref{sec:conclusion} with a discussion of the implications of this work on soft robotics applications.

\begin{figure*}[t]
	\centering
	\includegraphics[width=\textwidth, trim = {0pt 0pt 0pt 0pt}, clip = false]{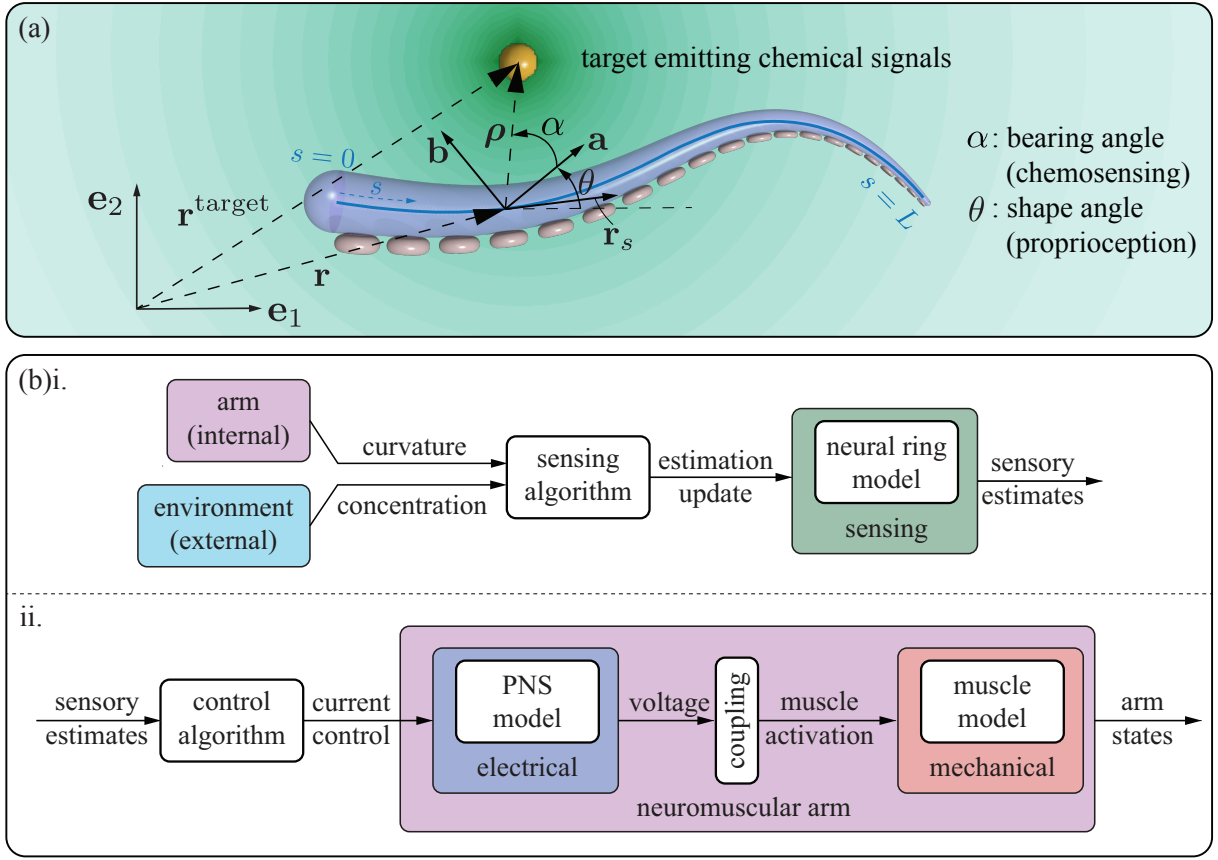}
	\vspace{-20pt}
	\caption{The sensorimotor control problem: (a) Chemical signals diffused from the target are sensed by the suckers, based on which the arm needs to catch the target; and (b) An abstract block diagram of the sensorimotor control problem: i. The sensing system takes concentration from the environment (external) and curvature from the arm system (internal) as inputs and produces sensory estimates under proposed sensing algorithm; ii. The neuromuscular arm system takes the sensory estimates from the sensing system and produces arm states under proposed control algorithm.}
	\label{fig:problem_statement}
	\vspace{-10pt}
\end{figure*}

\section{Mathematical formulation of the sensorimotor control problem and overall architecture}\label{sec:math_problem}

An octopus arm is modeled as a flexible planar rod (see Figure~\ref{fig:problem_statement}(a)) with rest length $L$ in an inertial laboratory frame $\set{{\mathbf{e}}_1,\mathbf{e}_2}$. The arc-length coordinate of the centre line of the arm is denoted by $s\in[0,L]$. Physically, the centre line runs through the ANC. Along with $s$, the second independent coordinate is time $t$. For notational ease, the partial derivatives
with respect to $s$ and $t$ are denoted as $(\cdot)_s :=\frac{\partial }{\partial s}$ and $(\cdot)_t :=\frac{\partial }{\partial t}$, respectively.  

At the material point $s$ along the centerline, the {\em kinematic pose} of the arm is defined by the position vector $\mathbf{r}(s,t)  \in \R^2$ and the angle $\theta(s,t) \in[0,2\pi)$.  The angle $\theta$ defines the material frame $\{\mathbf{a}, \mathbf{b}\}$, where $\mathbf{a} = \cos \theta \,\mathbf{e}_1 + \sin \theta \, \mathbf{e}_2$ and $\mathbf{b} = -\sin \theta \,\mathbf{e}_1 + \cos \theta \, \mathbf{e}_2$.
In particular, the kinematic equations are 
\begin{equation}
	\mathbf{r}_s = \bm{\nu}, \quad \theta_s = \kappa, ~~ s \in [0, L]
	\label{eq:kinematics}
\end{equation}
where $\kappa$ is referred to as the curvature and $\bm{\nu} := \nu_1 \mathbf{a} + \nu_2 \mathbf{b}$ has as its components the stretch ($\nu_1$) and the shear ($\nu_2$).  For an inextensible ($\nu_1 \equiv 1$) and an unshearable ($\nu_2 \equiv 0$) rod, the curvature determines the entire `shape' of the rod (simply by integrating~\eqref{eq:kinematics}).  For this reason, the angle $\theta$ is referred to as the {\em shape angle}.  We clarify that the rod is {\em not} assumed to be inextensible or unshearable in this paper.  In fact, extensibility is an important characteristic of an octopus arm.

Apart from the arm, the other object of interest is a target (food) in the environment. Mathematically, the target is modeled as a point source located at $\mathbf{r}^\target \in \R^2$ in a steady fluid medium.  The target creates a concentration field $c=c(\bm{\mathsf{r}}, t)$, which is modeled according to the Fick's second law~\cite{fick1855ueber} of diffusion:
\begin{equation}
	\frac{\partial c}{\partial t} = D \nabla^2 c
	+ \frac{2\pi D}{\mu} \delta(\bm{\mathsf{r}} - \mathbf{r}^\target),\quad \bm{\mathsf{r}} \in\R^2,\;\;t>0	
	\label{eq:concentration_diffusion}
\end{equation}
where $\nabla^2$ is the 2D Laplacian, $D$ is the diffusivity parameter, $\delta(\cdot)$ is the Dirac delta function, and ${\mu}>0$ is an intensity parameter.

Along the arm, at the material point $s$, the concentration is denoted as $c(s,t):=c(\mathbf{r}(s,t),t)$, and the distance and bearing to the target are denoted as $\rho(s,t)$ and $\alpha (s,t)$, respectively.  The defining relationships for these are as follows (see Figure~\ref{fig:problem_statement}(a)):
\begin{align}
	\rho (s,t) = \abs{\mathbf{r}^{\text{target}} - \mathbf{r}(s,t)}, \quad
	\mathsf{R}(\alpha(s,t)) \, \mathbf{a} = \frac{\mathbf{r}^{\text{target}} - \mathbf{r}(s,t)}{\abs{\mathbf{r}^{\text{target}} - \mathbf{r}(s,t)}}
	\label{eq:definition-dist-bearing}
\end{align} 
where $\mathsf{R}(\alpha) = \left[\begin{smallmatrix} \cos \alpha & -\sin \alpha \\ \sin \alpha & \cos \alpha \end{smallmatrix}\right]$ is the planar rotation matrix. 

\begin{figure*}[!t]
	\centering
	\includegraphics[width=\textwidth, trim = {0pt 0pt 0pt 0pt}, clip = false]{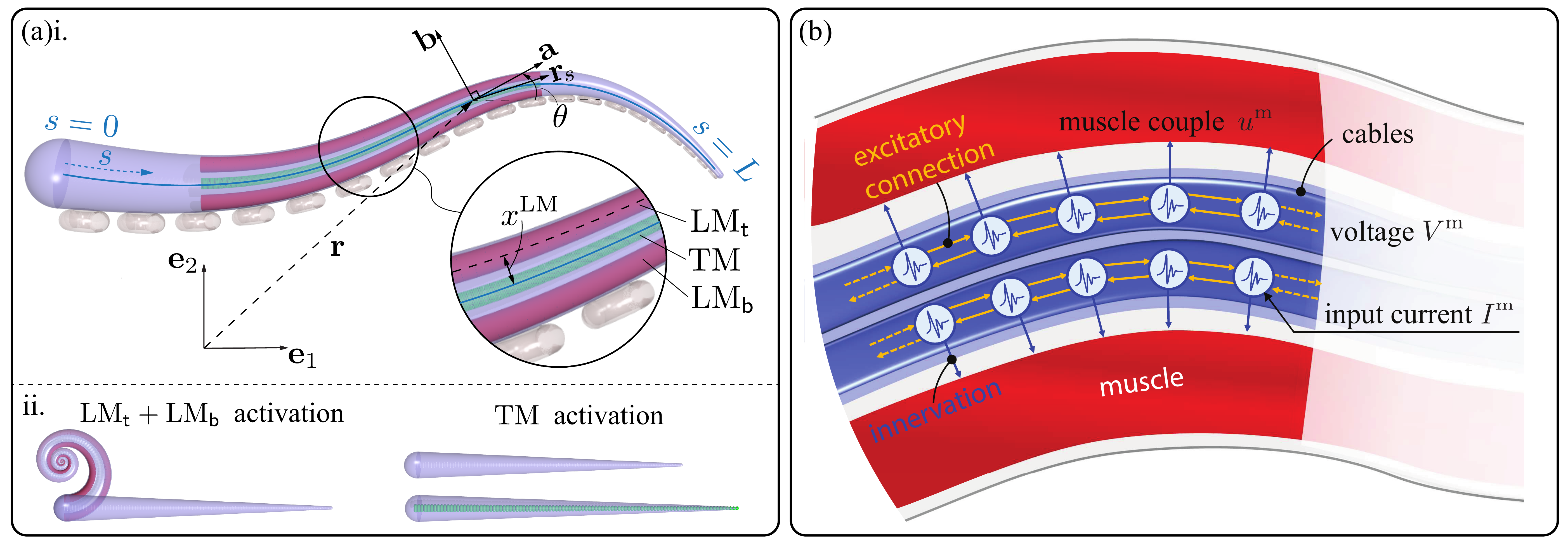}
	\vspace{-20pt}
	\caption{Modeling the neuromuscular control system of an octopus arm. (a)i. Cosserat rod model of a muscular soft arm is shown on the left. The musculature and sucker arrangements are shown in the inset. ii. Demonstrations of longitudinal muscles that bend and shorten the arm while the transverse muscle elongates the arm. (b) Cable-theoretic model of the arm PNS. Only the longitudinal muscles (in red) and their corresponding cables (in blue) are shown for clarity.}
	\label{fig:neuromuscular_model}
\end{figure*}

\medskip
In this paper, the sensorimotor control problem is divided into two types of problems -- control problem and sensing problem. Formulations of each of the problems with associated architectures are summarized next and described at length in the main body of this paper:

\smallskip
\noindent 
\textbf{(1) Control problem and architecture.} For planar motion of the arm, two types of muscles are modeled: (a) top and bottom longitudinal muscles, denoted as $\LMt$, and $\LMb$, respectively, and (b) transverse muscles, denoted as $\TM$ (see Figure~\ref{fig:neuromuscular_model}(a)). These three muscles suffice for motions of the arm in the plane. The control objective is to activate these muscles so that the arm reaches a target. This necessitates consideration of also the dynamics of the arm (in addition to the kinematics~\eqref{eq:kinematics}), as well as modeling of the PNS to provide neural activation of the muscles.  Figure~\ref{fig:problem_statement}(b)ii depicts the control architecture.

\noindent \textbf{(2) Sensing problem and architecture.}  Two types of sensory modalities are modeled: (a) chemosensors provide a measurement of the concentration $c(s,t)$ at discrete locations (suckers) along the arm, and (b) proprioceptors provide a measurement of the curvature $\kappa(s,t)$ at the sucker locations. The sensing objective is to fuse these sensory measurements to estimate the target location (bearing angle $\alpha(s,t)$ and distance $\rho(s,t)$) and the shape of the arm (angle $\theta(s,t)$). The former is an example of \textit{exogenous} or external sensing while the latter is an example of \textit{endogenous} or internal sensing. Figure~\ref{fig:problem_statement}(b)i depicts the architecture for modeling the external and internal sensing.

\medskip
While it is convenient to state these as separate problems, the two problems are coupled: control depends \textit{only} upon the information provided by the sensors while sensing is affected by the motion of the controlled arm.

\section{Control architecture} \label{sec:control_models}

The overall control architecture is depicted in Figure~\ref{fig:problem_statement}(b)ii and is comprised of models for the neuromuscular arm system (\S\,\ref{sec:control_neuro}), mathematical definition of the control problem and the proposed control law for the same (\S\,\ref{sec:control_problem}), as well as numerical simulations (\S\,\ref{sec:neuromuscular_results}). 

\subsection{Neuromusculature: arm, muscles, and the PNS}\label{sec:control_neuro}

The neuromuscular arm system has two main parts: (a) the flexible arm including its internal musculature, and (b) the peripheral nervous system (PNS) that includes the axial nerve cord (ANC).  

\paragraph{Geometric arrangement of the muscles.}  For a generic muscle $\muscle \in \mathcal{M} = \{\LMt, \LMb, \TM \}$, its relative position vector, with respect to the centerline, is $\bm{x}^\muscle = x^\muscle_1\, \mathbf{a} + x^\muscle_2 \, \mathbf{b}$. For the longitudinal muscles, the relative position vectors are $\bm{x}^\LMt = x^\LM \, \mathbf{b}$ and $\bm{x}^\LMb = -x^\LM \, \mathbf{b}$, where $x^\LM$ is the relative distance between the muscle and the centerline (see Figure~\ref{fig:neuromuscular_model}(a)).
The transverse muscles densely surround the centerline, and therefore we take their relative position to be zero, i.e. $\bm{x}^\TM = \bm{0}$.  
The organization of the musculature is portrayed in Figure~\ref{fig:neuromuscular_model}(a) and muscle related notations are tabulated in Table~\ref{tab:nomenclature_arm}. 

\paragraph{Modeling of the flexible arm with internal musculature} \hspace*{-8pt} $u^\muscle \longmapsto (\r,\theta)${\bf.} ~~ The flexible arm is modeled as a planar (2D) Cosserat rod~\cite{antman1995nonlinear, gazzola2018forward, chang2021controlling} 
\begin{align}
	\begin{bmatrix} (\varrho A \mathbf{r}_t)_t \\ (\varrho \mathrm{I} \theta_t)_t \end{bmatrix}
	= \underbrace{\begin{bmatrix} \mathbf{n}_s \\ m_s + (\nu_1 n_2 - \nu_2 n_1) \end{bmatrix}}_{\text{internal stresses}}
	-  \underbrace{\damping \begin{bmatrix} \mathbf{r}_t \\ \theta_t  \end{bmatrix}}_{\text{dissipation}} 
	+ \underbrace{\begin{bmatrix} \mathbf{f}^{\text{drag}} \\ 0 \end{bmatrix}}_{\text{fluidic drag}}
	\label{eq:dynamics}
\end{align}
Because the base of the arm ($s=0$) is attached to the head and the tip of the arm ($s=L$) is allowed to move freely, the boundary conditions are of the fixed-free type as follows:
\begin{align}
	\mathbf{r}(0, t) = \bm{0},~ \theta(0, t) = 0, ~\mathbf{n}(L, t) = \bm{0},~ m(L, t) = 0
	\label{eq:arm_boundary_conditions}
\end{align}
To complete the description of the arm elastodynamics, it remains to specify a model for the arm internal stresses $(\mathbf{n}, m)$, where $\mathbf{n} = n_1 \mathbf{a} + n_2 \mathbf{b}$. These are modeled as 
\begin{align*}
	\mathbf{n} = \mathbf{n}^{\text{e}} + \sum\limits_{\muscle \in \mathcal{M}} \mathbf{n}^\muscle, \quad m = m^{\text{e}} + \sum\limits_{\muscle\in \mathcal{M}} m^{\muscle} 
\end{align*} 
where $(\mathbf{n}^{\text{e}}, m^{\text{e}})$ is the passive component because of the elastic properties of the rod, and $(\mathbf{n}^\muscle, m^\muscle)$ is the active component because of the activation of the muscles. The model for passive elasticity of the arm $(\mathbf{n}^{\text{e}}, m^{\text{e}})$ is standard~\cite{chang2023energy} (see~\ref{appdx:neuromuscular}) while the model for the active muscle stresses is described next.

Each muscle $\muscle$ is activated through an {\em activation function} (input) $u^\muscle(s,t)$ which takes values in the interval $[0,1]$: $u^\muscle=0$ means that the muscle is inactive (and therefore produces no additional forces and couples) and $u^\muscle=1$ means that the muscle is maximally activated. Physically, $u^\muscle$ models the neural stimulation from the PNS, specifically the normalized firing frequency of the innervating motor neurons~\cite{nesher2020octopus}. In terms of $u^\muscle$, the active muscle force is expressed as
\begin{align}
	\mathbf{n}^\muscle (s,t) = u^\muscle (s,t) \musclecoeff^\muscle (s,t) \, \mathbf{t}^\muscle (s,t)
	\label{eq:muscle_force}
\end{align}
where the function $\musclecoeff^\muscle (s,t)$ accounts for the Hill-type modeling of the muscle force~\cite{hill1938heat, fung2013biomechanics} (details appear in~\ref{appdx:neuromuscular}) and $\mathbf{t}^\muscle$ is the direction along which the muscle force is applied.
For the two types of muscles, $\mathbf{t}^\muscle$ is as follows:
\begin{itemize}
	\item Contractions of longitudinal muscles cause shortening of the arm, thus the longitudinal muscle force is along the axial direction, i.e., $\mathbf{t}^\muscle =  \mathbf{a},~ \muscle \in \{\LMt, \LMb\}$.
	\item Transverse muscle contractions cause the arm to extend, therefore transverse muscle force is along the negative axial direction, i.e., $\mathbf{t}^{\TM} = - \mathbf{a}$.
\end{itemize}

Because the LMs are offset from the centerline, the active muscle force also results in a couple which is obtained by taking the cross product of the relative muscle position vector $\bm{x}^\muscle$ and the muscle force $\mathbf{n}^\muscle$ as follows:
\begin{align}
	m^\muscle = x^\muscle_1 n^\muscle_2 - x^\muscle_2 n^\muscle_1,\quad \muscle\in\mathcal{M}
	\label{eq:muscle_couple} 
\end{align}
Note that $\bm{x}^\TM = \mathbf{0}$, so the transverse muscles do not produce any couple on the arm. In summary, when activated, the longitudinal muscles serve to locally shorten and bend the arm, and the transverse muscles locally extend the arm, as illustrated in Figure~\ref{fig:neuromuscular_model}(a)ii. 

Table~\ref{tab:nomenclature_arm} tabulates the nomenclature for the various parameters related to the arm and musculature. Additional details of the arm passive elasticity, drag forces $\mathbf{f}^{\text{drag}}$, and muscle modeling are provided in~\ref{appdx:neuromuscular}.

\begin{table}[t]
	\footnotesize
	\centering
	\caption{Nomenclature -- Cosserat rod model of a muscular arm}
	\vspace{-8pt}
	\begin{tabular}{ll | ll | ll}
		\rowcolor{black}
		\multicolumn{4}{c}{\color{white} Arm related variables} & 
		\multicolumn{2}{c}{\color{white} Generic muscle related variables}\\
		$\r$ & center line position & $L$ & rest length of the arm & $\r^\muscle$ & muscle position\\
		$\theta$ & center line orientation & $\varrho$ & density of the arm & $\bm{x}^\muscle$ & muscle relative position\\
		$\nu$ & stretch and shear & $A$ & arm cross sectional area & $\mathbf{t}^\muscle$ & muscle tangent vector\\
		$\kappa$ & curvature  & $\mathrm{I}$ & second moment of area & $\ell^\muscle$ & muscle local length\\
		$\mathbf{n}$ & internal forces & $E$ & Young's modulus & $\mathbf{n}^\muscle$ & muscle force \\
		$m$ & internal couple & $G$ & shear modulus & $m^\muscle$ & muscle couple\\
		\hline
	\end{tabular}
	\label{tab:nomenclature_arm}
\end{table}

\paragraph{Modeling of the PNS} \hspace*{-8pt} $\Imuscle \longmapsto \Vmuscle$ {\bf.} ~~ Each of the three muscles -- $\LMt, \LMb$, and $\TM$ -- are independently controlled by the PNS. The central component of the PNS is the ANC, which is modeled by three long cylindrical nerves, referred to here as `cables' (see Figure~\ref{fig:neuromuscular_model}(b)). The neural activities of these nerves are modeled using cable theory which is a simplified model for one-dimensional excitable media~\cite{rall1962theory, tuckwell1988introduction}. 
For each muscle $\muscle \in \mathcal{M}$, the associated cable is described by the PDEs
\begin{subequations}
	\begin{align}
		\left(\tau { \Vmuscle} \right)_t &= \lambda^2 \Vmuscle_{ss} - \Vmuscle - \VmA + \Imuscle \label{eq:cable_eq_V} \\
		\left( \tauAdapt { \VmA}\right)_t &= -\VmA + bg(\Vmuscle) \label{eq:cable_eq_W}
	\end{align}
	\label{eq:cable_eq_full}
\end{subequations}
where $\Vmuscle = \Vmuscle (s,t)$ is the membrane potential, $\VmA = \VmA (s,t)$ is the adaptation variable to model self-inhibition~\cite{matsuoka1984dynamic}, and $\Imuscle = \Imuscle (s,t)$ is the stimulus current (control) which is produced by the local neural circuitry. The cable equation for voltage $\Vmuscle$ is accompanied by appropriate boundary conditions. Table~\ref{tab:nomenclature_cable}  tabulates the nomenclature for the various parameters related to the PNS.

\begin{table}[!t]
	\footnotesize
	\centering
	\caption{Nomenclature -- neural motor control model}
	\vspace{-8pt}
	\begin{tabular}{ll | ll }
		\rowcolor{black}
		\multicolumn{4}{c}{\color{white} Arm PNS related variables} \\
		$\Vmuscle$  & cable voltage 			    &  $\tau$ 			& time constant\\
		$\VmA$        & adaptation variable  	    &  $\tauAdapt$	& recovery rate \\
		$\Imuscle$  & total stimulus current  & $\lambda$ 		& length constant \\
		$b$    			 & adaptation parameter  & $g(\cdot)$ 		&  neuronal output function (ReLU) \\
		\hline
	\end{tabular}
	\label{tab:nomenclature_cable}
\end{table}

\paragraph{Neuromuscular coupling} \hspace*{-8pt} $\Vmuscle \longmapsto u^\muscle$ {\bf.} ~~
The muscle activation is modeled as a static function of the membrane potential as follows:
\begin{equation}
	u^\muscle (s,t) = \sigma(V^\muscle (s,t))
	\label{eq:neuromuscular_mapping}
\end{equation}
where $\sigma(\cdot)$ is a saturation-like function whose explicit formula appears in~\ref{appdx:numerics_setup}. Such a model is inspired by prior studies that suggest an approximately linear transformation of motor neuronal activity into muscular activation~\cite{matzner2000neuromuscular,nesher2020octopus}.  Similar models have been considered  for modeling skeletal muscles~\cite{hatze1977myocybernetic, audu1985influence}.

\medskip
Before describing the control problem, we first describe the equilibrium or the rest shapes of the arm obtained using the model.

\begin{figure*}[t]
	\centering
	\includegraphics[width=\textwidth, trim = {0pt 0pt 0pt 0pt}]{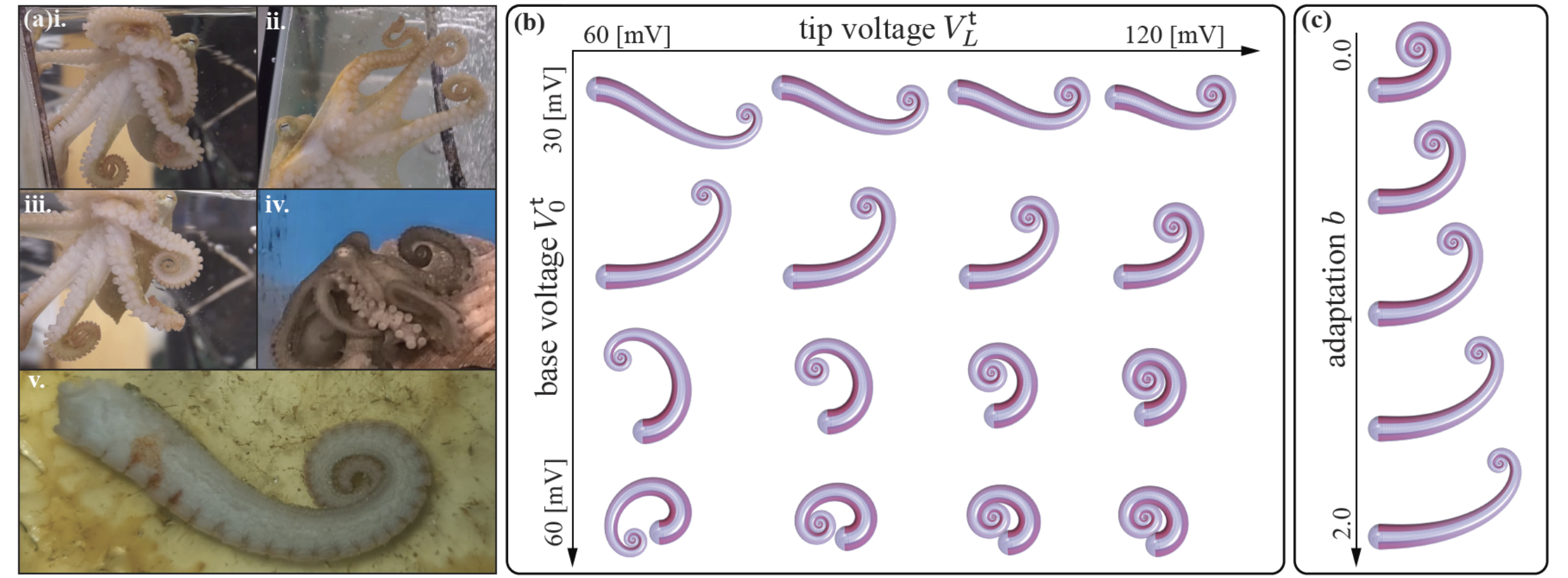}
	\vspace{-20pt}
	\caption{Curled rest shapes in both real octopus arms and simulated arms. (a) Snapshots of different octopuses showing curled arm at rest. Subfigures i-iv are of live octopuses while the subfigure v is an severed arm that remains curled even after isolation. (b) The equilibrium of the arm under varying base and tip voltages $\Vtop_0$ and $\Vtop_L$ for the top longitudinal muscle. The intensities of the distributed muscle actuation are illustrated in red. (c) The equilibrium of the arm under varying adaptation parameter $b$.}
	\label{fig:curl}
	\vspace{-10pt}
\end{figure*}

\paragraph{Biological observations of rest shape.}
At rest, octopus arms tend to curl and form spirals, exposing suckers outwardly (see Figure~\ref{fig:curl}(a)). This curling behavior may be beneficial for several biophysical reasons: for protecting the arm from predators, for increased environmental awareness owing to a large number of exposed suckers, or for providing an efficient initial state for arm reaching motions~\cite{packard1971body, mather1998octopuses}.
Notably, curling at rest seems to persist even without active input from the central nervous system (CNS), as observed in isolated arms (Figure~\ref{fig:curl}(a)v). Mechanically, arm curling at rest may involve inherent tensions of the arm muscle groups~\cite{di2021beyond}.

\paragraph{Numerical results.} 
The rest shapes are obtained by setting $\Imuscle = 0$ and specifying a fixed-fixed boundary condition 
\begin{align}
	\Vmuscle (0) = \Vmuscle_0, ~~ \Vmuscle (L) = \Vmuscle_L
	\label{eq:cable_eq_bc_rest}
\end{align}
where $\Vmuscle_0$ and $\Vmuscle_L$ are the fixed boundary voltages that need to be specified.
As a function of the boundary voltages $\Vmuscle_0$ and $\Vmuscle_L$, explicit formulae of the rest state voltage profiles $V^\muscle$ are provided in~\ref{appdx:rest_state} as Lemma~\ref{lemma:equilibrium_no_control}.
To illustrate the rest shapes, an inextensible and unshearable arm is considered for simplicity. The adaptation parameter $b = 1.0$ in~\eqref{eq:cable_eq_W}, and the boundary voltages $\Vbot_0=40$ [mV] and $\Vbot_L=0$ [mV] for the bottom longitudinal muscle in~\eqref{eq:cable_eq_bc_rest} are chosen. The boundary voltages of the top longitudinal muscle $\Vtop_0$ and $\Vtop_L$ are taken from the sets $\{30,40,50,60\}$ [mV] and $\{60,80,100,120\}$ [mV], respectively. This results in a 4-by-4 chart as illustrated in Figure~\ref{fig:curl}(b).
It is observed that as $\Vtop_0$ increases (columns), the arm tends to bend more at the base. On the other hand, as $\Vtop_L$ increases (rows), the tip curls more. 
Moreover, the shape in row-2/column-4 qualitatively matches Figure~\ref{fig:curl}(a)i-iii, where the arm is straight near the base and curls up at the tip. Similarly, the row-4/column-2 result shows a tip curling with a bent base, matching the recording Figure~\ref{fig:curl}(a)iv. Finally, row-1/column-4 presents a shape similar to the isolated curled arm in Figure~\ref{fig:curl}(a)v. 

In part (c) of Figure~\ref{fig:curl}, the effect of varying the adaptation parameter $b$ is shown. In these solutions, the boundary voltages of the top longitudinal muscle are fixed at $\Vtop_0=40$ [mV] and $\Vtop_L=80$ [mV]. The parameter $b$ is taken from the set $\{0,0.5,1.0,1.5,2.0\}$. It is seen that as the adaptation increases, the arm tends to unwind, or in other words, the arm is less curled.

\subsection{Control problem and its solution}\label{sec:control_problem}

\paragraph{Statement of the control problem.}
For each muscle $\muscle \in \mathcal{M}$, the control input is the current $I^\muscle (s,t)$.  The control problem is to design these inputs such that the arm reaches a fixed point target at the location $\mathbf{r}^{\text{target}}$. For this purpose, the following definition is introduced.

\begin{definition} {\bf ($\epsilon$-reach)}
	An arm is said to $\epsilon$-reach a target if there exists an arc-length $s^* \in [0, L]$ and a time $T \in [0, \infty)$, such that $\rho (s^*, t) \leq \epsilon$ for all $t \geq T$.
	\label{def:epsilon_reach}
\end{definition}

\begin{figure*}[t]
	\centering
	\includegraphics[width=\textwidth, trim = {0pt 0pt 0pt 0pt}, clip=false]{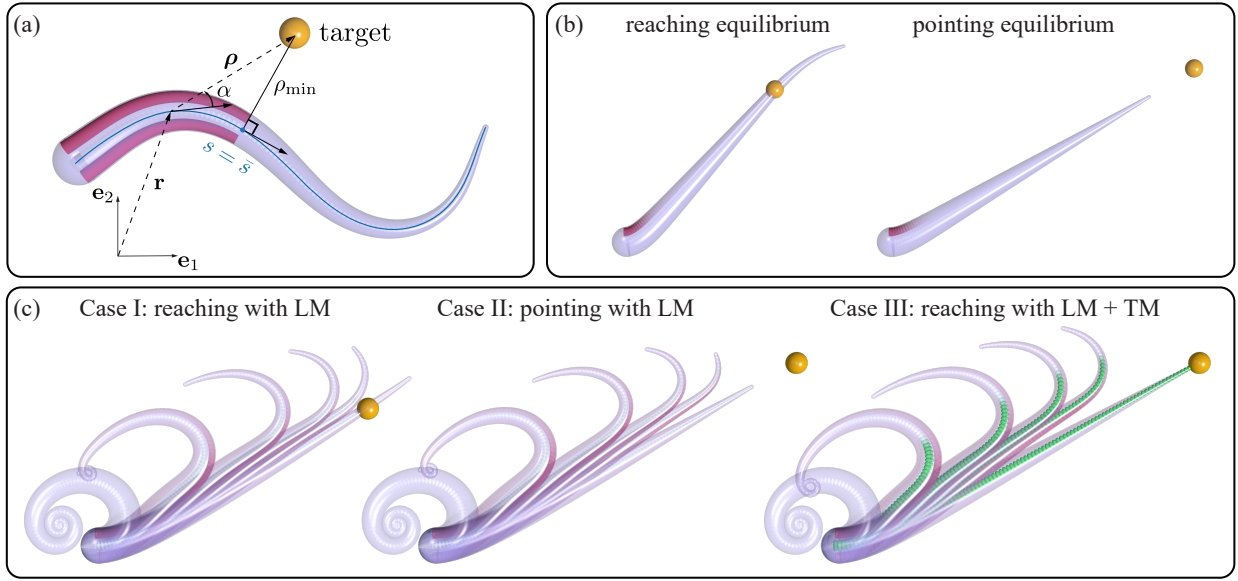}
	\vspace{-20pt}
	\caption{(a) The schematic of the neuromuscular control that uses the sensory information of bearing $\alpha$ and the arc-length of the closest point to the target $s=\bar{s}$. (b) Two equilibrium configurations: \textit{reaching} and the \textit{pointing}.
		(c) Neuromuscular control in octopus arm. Three cases are presented. The rods are initialized with a bend at the base and the curl at the tip. The red color represents both top and bottom longitudinal muscle activations. Transverse muscle activation is in green. Case I: static target is within the reach of the rod and only the longitudinal muscles are activated. The rod does bend propagation towards the target under the neuromuscular control and finally stabilizes at the \textit{reaching} equilibrium. Case II: a static target is outside of the reach. Only the longitudinal muscles are activated. The rod eventually stabilizes at the \textit{pointing} equilibrium. Case III: a static target is outside of the reach. All three muscles (including the transverse muscle) are activated. The rod elongates and stabilizes at the equilibrium which reaches the target.}
	\label{fig:reaching}
	\vspace{-10pt}
\end{figure*}

\paragraph{Proposed feedback control law.} Denote the point on the arm closest to the target as follows:
\begin{align}
	\bar{s}(t) := \argmin\limits_{s \in [0, L]}  ~ \rho (s,t)
	\label{eq:definition-sbar}
\end{align}
In terms of the bearing $\alpha(s,t)$ and $\bar{s}(t)$, the proposed current controls for the three muscles are as follows:
\begin{subequations}
	\begin{align}
		I^{\LMt} (s,t) = f^{\LMt} (\alpha(s,t), \bar{s}(t)) &:= \begin{cases}
			\chi \sin (\alpha(s,t)) \1 \{ s \leq \bar{s}(t) \}, \hspace{0.22in} \text{if}~ \sin (\alpha (s,t)) \geq 0 \\
			0, \hspace{1.48in} \text{otherwise}
		\end{cases} \label{eq:control_LM_top}
		\\
		I^{\LMb} (s,t) = f^{\LMb} (\alpha(s,t), \bar{s}(t)) &:= \begin{cases}
			-\chi \sin (\alpha(s,t)) \1 \{ s \leq \bar{s}(t) \}, \quad \text{if}~ \sin (\alpha (s,t)) \leq 0 \\
			0, \hspace{1.5in} \text{otherwise}
		\end{cases} \label{eq:control_LM_bottom}
		\\
		I^{\TM} (s,t) = f^{\TM} (\alpha(s,t), \bar{s}(t)) &:= \chi \cos^2(\alpha(s,t))\mathds{1}\{s\leq\bar{s} (t)\}, \label{eq:control_TM}
	\end{align}
	\label{eq:sensory_feedback_control_law}
\end{subequations}
where $\chi$ is a positive constant and $\mathds{1}\{ \cdot \}$ denotes the indicator function. In words, the control law~\eqref{eq:sensory_feedback_control_law} makes the arm active up to the $\bar{s}$ point ($0\leq s \leq \bar{s}(t)$), and passive beyond it ($\bar{s}(t) \leq s \leq L$). Figure~\ref{fig:reaching}(a) illustrates all variables regarding the control law. 

\begin{remark} {\it \bf (Rationale for the control law)} \label{remark:motion_camouflage}
	The proposed control law~\eqref{eq:sensory_feedback_control_law} is inspired from two bodies of literature:
	
	(i) The use of the indicator function is based upon experimental findings in octopus arms. Specifically, electromyogram (EMG) recordings suggest that during bend propagation maneuver, the arm is passive beyond the bend point~\cite{gutfreund1998patterns, sumbre2001control}. A traveling wave ansatz for open-loop control of bend propagation was first considered in~\cite{yekutieli2005dynamic} for a discrete mass-spring model and in~\cite{wang2022control} for a continuum model of a soft arm.  
	
	(ii) The form of the control law involving $\sin (\cdot)$ of the bearing is inspired by the models of the motion camouflage steering strategy~\cite{glendinning2004mathematics, justh2006steering} that is observed in many natural predators, including bats and falcons. 
	In particular, the arm equilibrium spatial configuration has parallels with the temporal trajectory of a pursuer intercepting a prey under the motion camouflage strategy, as discussed in~\ref{appdx:motion_camouflage}. 
	It is shown that for $\nu_1 \equiv 1$ and $\nu_2 \equiv 0$, the arm equilibrium geometry~\eqref{eq:sensory_kinematics_full} and corresponding control laws~(\ref{eq:control_LM_top},\, \ref{eq:control_LM_bottom}) can be viewed as parallels to the time evolution of the motion camouflage trajectory~\eqref{eq:bearing_dynamics} and its corresponding steering control~\eqref{eq:ctrl_law_MC}, respectively. 
\end{remark}

\paragraph{Equilibrium and its analysis.}
An equilibrium of the arm is obtained by equating the left-hand sides of the dynamics of the neuromuscular arm~ (\ref{eq:dynamics},\,\ref{eq:cable_eq_full},\,\ref{eq:neuromuscular_mapping}) to zero, with the consideration of the boundary conditions (\ref{eq:arm_boundary_conditions}) and the feedback control law \eqref{eq:sensory_feedback_control_law}. This yields the following set of equations:

\smallskip
\noindent
(i) kinematics of the arm as given by equation~\eqref{eq:kinematics};

\noindent
(ii) statics of the rod (balance of passive and active stresses) from equilibrium analysis of~\eqref{eq:dynamics} and boundary conditions~\eqref{eq:arm_boundary_conditions}
\begin{align}
	\begin{split}
		\mathbf{n}^{\text{e}} (s, (\nu_1, \nu_2, \kappa)) &= - \sum\limits_\muscle \mathbf{n}^\muscle (s, (\nu_1, \nu_2, \kappa); u^\muscle(s)), \\
		m^{\text{e}} (s, (\nu_1, \nu_2, \kappa)) &= - \sum\limits_\muscle {m}^\muscle (s, (\nu_1, \nu_2, \kappa); u^\muscle(s)); 
	\end{split}
	\label{eq:statics_arm}
\end{align}
(iii) the neuromuscular coupling $u^\muscle = \sigma (V^\muscle)$ given by equation~\eqref{eq:neuromuscular_mapping};

\noindent
(iv) statics of the cables~\eqref{eq:cable_eq_full} with a free-free boundary condition
\begin{align}
	\lambda^2 V^\muscle_{ss} - V^\muscle - b g(V^\muscle) + I^\muscle = 0, ~~ V^\muscle_s (0) = V^\muscle_s(L) = 0;
	\label{eq:statics_cable} 
\end{align}

\noindent
(v) the relative geometry of the arm and the target, obtained by taking spatial derivatives of the definitions of ($\rho$, $\alpha$)~\eqref{eq:definition-dist-bearing}, and using the kinematics~\eqref{eq:kinematics}
\begin{equation}
	\begin{aligned}
		\dist_s &= -( \nu_1 \cos \alpha + \nu_2 \sin \alpha), \quad \rho(0) = \abs{\r^{\text{target}}}  \\
		\alpha_s &= -\kappa + \frac{1}{\dist} (\nu_1 \sin \alpha - \nu_2 \cos \alpha), \quad \mathsf{R}(\alpha (0))\, \mathbf{e}_1 = \frac{\r^{\text{target}}}{\rho(0)},
	\end{aligned}
	\label{eq:sensory_kinematics_full}
\end{equation} 
where $(\rho(0), \alpha(0))$ are calculated from the definition~\eqref{eq:definition-dist-bearing} and the arm's fixed boundary condition at $s=0$ as given in~\eqref{eq:arm_boundary_conditions}.

The equilibrium is computed using an iterative numerical procedure as follows.
Notice first that the deformations of the arm  $(\nu_1, \nu_2, \kappa)$ uniquely determine the equilibrium configuration $(\r, \theta)$ of the arm by integrating the kinematics~\eqref{eq:kinematics}. 
A numerical iterative approach is taken to obtain the deformations with the following steps: (a) starting with some choice of $(\nu_1, \nu_2, \kappa)$ and a given  target location, the equations \eqref{eq:sensory_kinematics_full} are first solved for $(\rho, \alpha)$ which yields the current inputs $I^\muscle$ following \eqref{eq:sensory_feedback_control_law}; (b) the cable static equation~\eqref{eq:statics_cable} is solved next, resulting in the static voltages $V^\muscle$; (c) static muscle activations $u^\muscle$ are computed using the neuromuscular coupling~\eqref{eq:neuromuscular_mapping}; and finally (d) the two sides of the rod statics~\eqref{eq:statics_arm} are compared to generate the required iterative updates for the deformations $(\nu_1, \nu_2, \kappa)$.

The equilibrium is denoted using a superscript $^{\star}$. Specifically, $(\rho^\star(s), \alpha^\star(s)), ~ s\in [0, L]$ is the distance and bearing to the target, and $\bar{s}^\star$ is the arc-length of the closest point to the target. The main results about the equilibrium and its dynamic stability are described by the following theorem.

\begin{theorem}
	Suppose the arm is inextensible ($\nu_1 \equiv 1$) and unshearable ($\nu_2 \equiv 0$). Then,
	
	\smallskip
	\noindent
	A. {\bf (Equilibrium analysis)} For any given $\epsilon >0$, there exists a $\bar{\chi} = \bar{\chi} (\epsilon)$, such that for all $\chi > \bar{\chi}$
	
	(a) if $\rho^\star (0) \leq L$, then $\rho^\star(\bar{s}^\star) \leq \epsilon$,
	
	(b) if $\rho^\star (0) > L$, then $\cos (\alpha^\star (L)) \geq 1 - \epsilon$.
	
	\smallskip
	\noindent
	B. {\bf (Dynamic stability)} For such a choice of $\chi$, the equilibrium is locally asymptotically stable. 
	\label{thm:sensoryfeedback_control}
\end{theorem}

\begin{proof}
	See~\ref{appdx:sensoryfeedback_control_proof}. 
\end{proof}

Figure~\ref{fig:reaching}(b) illustrates the two cases for the equilibrium characterized in Theorem~\ref{thm:sensoryfeedback_control}-A: 

\textbf{(i) Reaching equilibrium:} If $\rho^\star (0) \leq L$, the target is in the reachable workspace of an inextensible arm. In this case, $\rho^\star (\bar{s}^\star) \leq \epsilon$ means that the arm $\epsilon$-reaches the target (in the sense of Definition~\ref{def:epsilon_reach}). 

\textbf{(ii) Pointing equilibrium:} If $\rho^\star(0) >L$, the target is outside the reachable workspace. In this case, $\cos (\alpha^\star (L)) \geq 1 - \epsilon$ means that the bearing at the tip of the arm becomes arbitrarily small. This implies that the arm points towards the target.

\subsection{Simulation results for neuromuscular control} \label{sec:neuromuscular_results}

To illustrate the two different cases (reaching and pointing) in Theorem~\ref{thm:sensoryfeedback_control}-A, two sets of numerical experiments are carried out for an inextensible and unshearable arm. Under these conditions, it suffices to activate only the longitudinal muscles. A third experiment is then performed to demonstrate the capabilities of the transverse muscles.
For all the experiments, the arm is initialized at the equilibrium configuration that corresponds to row-4/column-2 shape in Figure~\ref{fig:curl}(b). The numerical parameters are tabulated in Table~\ref{tab:parameters_neuromuscular}.

\smallskip
\noindent
\textbf{(1) Case I (reaching with LM):} For the first experiment, a static target is presented at the location $(0.75L, 0.375L)$. As this target is within the reach of the arm ($\dist(0)\leq L$), the arm reaches the target by using our prescribed feedback control. Moreover, as shown by the temporal snapshots, the arm maintains the initial bend and propagates it toward the tip. The arm stabilizes in a configuration which effectively \textit{reaches} the target. 

\smallskip
\noindent
\textbf{(2) Case II (pointing with LM):} For the second experiment, the target location is changed to $(1.0L, 0.5L)$, keeping all other simulation conditions the same as Case I. Such a placement of the target makes $\dist(0)>L$, i.e., outside the reach of the arm. As is seen from Figure~\ref{fig:reaching}(c), the arm dynamically stabilizes to a configuration that \textit{points} towards the target. 

\smallskip
\noindent
\textbf{(3) Case III (reaching with LM + TM):} Finally, to showcase the effect of the transverse muscles, the constraints of inextensibility and unshearability are dropped. All other simulation conditions are kept the same as the pointing case (Case II).
It is clearly seen from Figure~\ref{fig:reaching}(c) that the arm is able to extend and reach the target with the help of transverse muscle actuation (shown in green). The simulation also helps verify the dynamic stability of the feedback control system in Theorem~\ref{thm:sensoryfeedback_control}-B.

\smallskip
The proposed feedback control law~\eqref{eq:sensory_feedback_control_law} depends upon $(\alpha(s,t),\bar{s}(t))$. In a sensorimotor control setting, these are estimated from sensory measurments. The algorithm for the same is the subject of the next section.

\begin{table*}[!t]
	\footnotesize
	\centering
	\caption{Parameters for neuromuscular arm and control}
	\hspace*{-5pt}
	\begin{tabular}{ccc|ccc}
		\rowcolor{black}
		{\color{white} Parameter} & 
		{\color{white} Description} & 
		{\color{white} Value} & 
		{\color{white} Parameter} & 
		{\color{white} Description} & 
		{\color{white} Value} \\
		&&&&& \\ [-9pt]
		& {\bf Muscular arm model} & &
		& {\bf Initial arm configuration} & \\	
		$L$ & rest length of the arm [cm] & $20$ & 
		$\Vtop_0$ & base voltage for $\LMt$ [mV] & $60$ \\
		&&&&& \\ [-9pt]
		$\rodbase$ & rod base radius [cm] & $1$ &
		$\Vtop_L$ & tip voltage for $\LMt$ [mV] & $80$ \\
		&&&&& \\ [-9pt]
		$\rodtip$ & rod tip radius [cm] & $0.1$ &
		$\Vbot_0$ & base voltage for $\LMb$ [mV] & $40$ \\
		&&&&& \\ [-9pt]
		$\varrho$ & density [kg/${\text{m}}^3$] & $1042$ &
		$\Vbot_L$ & tip voltage for $\LMb$ [mV] & $0$ {\smallskip}\\
		\cline{4-6}\noalign{\smallskip}
		$\xi$ & damping coefficient [kg/s]  & $0.01$ &
		& {\bf Cable equation} & \\
		$E$ & Young's modulus [kPa] & $10$ &
		$\tau$ & time constant [s] & $0.04$ \\
		$G$ & shear modulus [kPa] & $10/3$ &
		$\tauAdapt$ & adaptation rate [s] & $0.4$ \\
		\cline{1-3}\noalign{\smallskip}
		& {\bf Drag model} & &
		$\lambda$ & length constant [cm] & $2$ \\
		$\rhow$ & water density [kg/${\text{m}}^3$] & $1022$ &
		$b$ & adaptation parameter & $1$ \\
		\cline{4-6}\noalign{\smallskip}
		$\upxi\mytan$ & tangential drag coefficient & $0.155$ &
		& {\bf Neuromuscular control} & \\
		$\upxi\myper$ & normal drag coefficient & $5.065$ &
		$\chi$ & control gain & $200$ \\
		\hline
	\end{tabular}
	\label{tab:parameters_neuromuscular}
\end{table*}

\section{Sensing architecture}\label{sec:sensing_models}

The overall sensing architecture is depicted in Figure~\ref{fig:problem_statement}(b)i. A central component of the sensing neuroanatomy is the model of the neural ring which is used to encode an angle variable. The model is presented in \S\,\ref{sec:sensing_neuro}, which is followed by the mathematical definition of the sensing problem (\S\,\ref{sec:sensing_problem}), proposed sensing algorithm and its analysis (\S\,\ref{sec:algorithm_sensing}), as well as simulation results (\S\,\ref{sec:sensing_results}). 

\subsection{Neuroanatomy: neural rings and sensing units}\label{sec:sensing_neuro}

\begin{figure*}[!t]
	\centering
	\includegraphics[width=\textwidth, trim = {0pt 0pt 0pt 0pt}, clip = false]{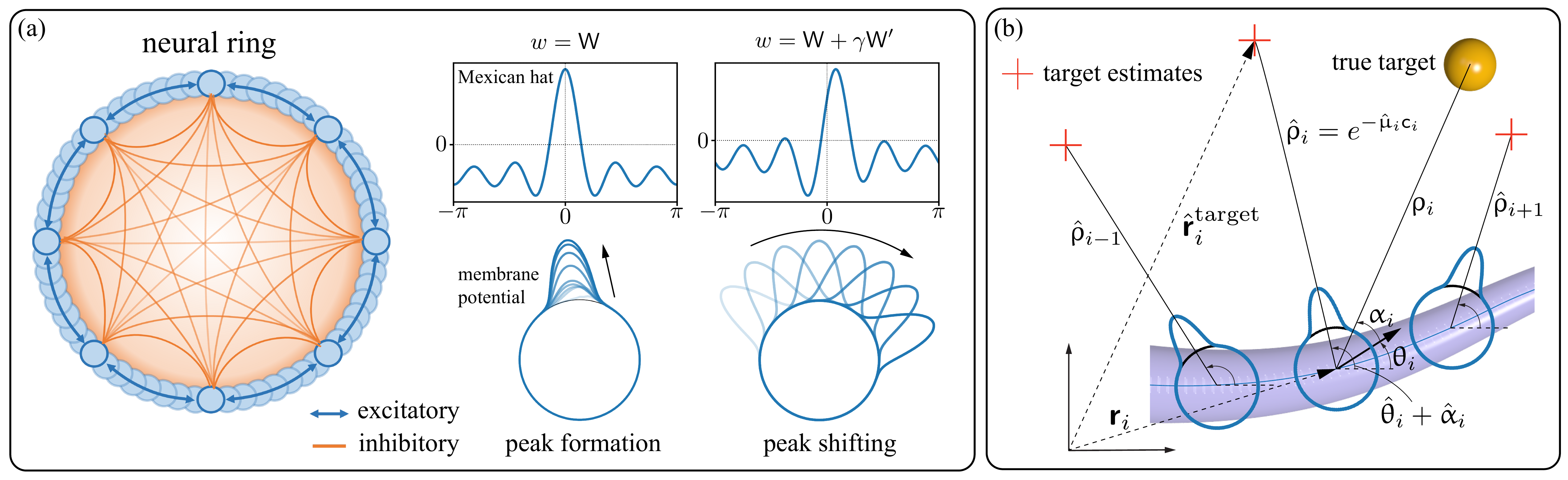}
	\vspace{-20pt}
	\caption{Modeling the sensing system. (a) Neural ring architecture and its properties. The weight function without shift is a Mexican hat function where its positive part creates excitatory connections between neurons while the negative part corresponds to inhibition. Without shift, the membrane potential forms a peak. With shift $\gamma$, the peak shifts with speed that depends on $\gamma$. (b) Sensing as a consensus problem. Each sensor has its own estimates of orientation $\hat{\uptheta}_i$, target bearing $\hat{\upalpha}_i$, and the concentration intensity parameter $\hat{\upmu}_i$ for $i=1,\ldots,N$ which create a target estimate $\rt_i$ (red plus signs). The goal is for all sensors to reach a consensus on their target beliefs and eventually estimate the true target location.}
	\label{fig:sensing_model}
	\vspace{-10pt}
\end{figure*}

\paragraph{Neural ring.}
A neural ring is a one-dimensional continuum of a fully-connected network of neurons (see Figure~\ref{fig:sensing_model}(a))~\cite{amari1977dynamics, zhang1996representation, hahnloser2003emergence}. 
The membrane potential of a neural ring is denoted by $\mathsf{V}(\varphi,t)$ where $\varphi\in[0,2\pi)$.  
The evolution of the membrane potential is according to the  integro-differential equation~\cite{zhang1996representation}
\begin{equation}
	\tauRing \frac{\partial }{\partial t} \mathsf{V} (\varphi,t) = -\mathsf{V} (\varphi,t) + \int_{0}^{2\pi} w(\varphi-\phi,t) h \left(\mathsf{V}(\phi,t)\right)\ud \phi 
	\label{eq:neural_ring}
\end{equation}
where $w(\cdot)$ and $h(\cdot)$ are referred to as the weight function and the synaptic response function, respectively.  In this paper, the weight function is assumed to be of the form   
\begin{equation}
	w(\varphi,t) = \mathsf{W} (\varphi) + \gamma(t) \mathsf{W}^\prime(\varphi)
	\label{eq:weight_function}
\end{equation}
where $\mathsf{W} (\cdot)$ is a periodic function of its argument, $\mathsf{W}^\prime (\cdot)$ is its derivative, and $\gamma(t)$ is the input (that needs to be designed).  The function $\mathsf{W}$ is taken to be the Mexican hat function as plotted in Figure~\ref{fig:sensing_model}(a)~(see~\ref{appdx:weight} and \cite{zhang1996representation} for details).  Such a choice of the weight function models the local excitation (for positive values of the weight) and global inhibition (for negative values of the weight). This phenomenon is called local autocatalysis (excitation) with lateral inhibition (LALI)~\cite{boettiger2009neural, murray2003introduction}.
Two important properties of the neural ring are as follows~\cite{zhang1996representation}: 
\begin{itemize}
	\item $\left[\gamma(t) = 0\right]$
	When $\gamma(t) = 0$, the equilibrium of the membrane voltage profile has a single maximum (see Figure~\ref{fig:sensing_model}(a)).  There is a rotational symmetry whereby the equilibrium can be rotated to yield another equilibrium.  Therefore, the maximum may form at any angle in $[0,2\pi)$. Denote such an equilibrium voltage profile by $\bar{\mathsf{V}}(\cdot)$.
	\item $\left[\gamma(t) \neq 0\right]$
	Consider the neural ring~\eqref{eq:neural_ring} with the initial condition $\mathsf{V} (\varphi, 0) = \bar{\mathsf{V}}(\varphi)$. Then it is easy to verify that the following is a solution to~\eqref{eq:neural_ring}
	\begin{align*}
		\mathsf{V}(\varphi, t) = \bar{\mathsf{V}}\left(\varphi + \frac{1}{\tauRing} \int_0^t \gamma (t') dt' \right),~~ t> 0
	\end{align*}
	In words, a non-zero input $\gamma(t)$ causes a rigid dynamic rotation of the voltage profile $\bar{\mathsf{V}}(\cdot)$ with an instantaneous rotational speed $-\gamma(t)/\tauRing$. Next, define ${\varphi}^\star (t) := \argmax\limits_{\varphi \in [0,2\pi)} \mathsf{V}(\varphi, t)$. It then follows that 
	\begin{align}
		\frac{\dif }{\dif t} {\varphi}^\star (t) = -\gamma(t)/\tauRing, ~~ t>0
		\label{eq:angular_velocity-shifting-mapping}
	\end{align}
	See \cite[Appendix 4]{zhang1996representation} for a proof of this property. 
\end{itemize}

\begin{remark} {\bf (Rationale for neural rings)} \label{remark:neural_ring}
	Directional information is often encoded by ring-type neural structures in nature, e.g. the head direction cells (HD cells) in mice~\cite{taube1990head, taube1995head, ajabi2023population} and compass neurons within the central complex in flies~\cite{giraldo2018sun, fisher2019sensorimotor}. Essentially, this population of cells serve as a compass for the organism to orient themselves and navigate. Similar class of mathematical model of ring attractor networks has also been considered to model path integration, place cells, and grid cells~\cite{mcnaughton2006path, moser2017spatial, gardner2022toroidal}. We re-emphasize that we make no claim that such ring structures are encountered in the octopus. Thus, in lieu of the unknown sensing architecture of the octopus, we are inspired by the above biological systems and their robust performance.
\end{remark}

\vspace*{-10pt}
\paragraph{Geometric arrangement of sensing units.} $N$ sensing units are located along the length of the arm. Physically, the $i$-th sensing unit is at the location of the $i$-th sucker. Their locations are denoted as ${s}_i, ~i=1,2,..., N$, with ${s}_1 = 0$ (at the base), ${s}_N=L$ (at the tip), and $s_i \leq s_{i+1}$ for $i=1,2,..., N-1$. 
The arc-length difference between two sensing units is denoted by $\Delta s_{ij} := s_i - s_j$.
For the $i$-th sensing unit, its neighborhood $\mathcal{N}_i$ is the set of adjacent sensing units, i.e. $\mathcal{N}_i = \{ i-1, i+1\}, ~i=2,...,N-1$, and $\mathcal{N}_1 = \{2\}, ~\mathcal{N}_N = \{N-1\}$. 
The position and orientation of the $i$-th sensing unit (in the lab frame) are denoted as $\rsucker_i (t) := \mathbf{r}(s_i, t)$ and~${\uptheta}_i(t) := \theta(s_i, t)$, respectively. The true distance and bearing to target (in the material frame) are denoted as $\uprho_i (t) := \dist(s_i, t)$ and~$\upalpha_i := \alpha(s_i, t)$, respectively. See equation~\eqref{eq:definition-dist-bearing} for defining relationship, Figure~\ref{fig:sensing_model}(b) for illustration, and Table~\ref{tab:nomenclature_sensing} for a complete nomenclature.

\subsection{Statement of the sensing problem} \label{sec:sensing_problem}

\paragraph{Sensory inputs.} The sensory information available to the $i$-th sensing unit are the chemical concentration $\c_i (t) := c(s_i, t)$ from the chemosensors and the curvature $\upkappa_i(t) :=\kappa(s_i,t)$ from the proprioceptors.

\vspace*{-10pt}
\paragraph{Encoding of external and internal sensing.}  The $i$-th sensing unit is equipped with two neural rings to encode the estimates for the two angles $\uptheta_i (t)$ and $\upalpha_i(t)$. The voltages of the two neural rings are denoted as $\mathsf{V}^{\uptheta_i} (\cdot, t)$ and $\mathsf{V}^{\upalpha_i} (\cdot, t)$, while respective inputs to the rings are denoted as $\gamma^{\uptheta_i} (t)$ and $\gamma^{\upalpha_i} (t)$. In terms of the voltages, the estimates are defined as follows:
\begin{align}
	\hat{\uptheta}_i (t)  = \argmax\limits_{\varphi \in [0, 2\pi)}~ \mathsf{V}^{\uptheta_i}(\varphi, t), \qquad \hat{\upalpha}_i (t) = \argmax\limits_{\varphi \in [0, 2\pi)}~ \mathsf{V}^{\upalpha_i}(\varphi, t) 
	\label{eq:sucker_estimates}
\end{align}
In addition to the two angles, each sensing unit also maintains an estimate $\hat{\upmu}_i (t)$  of the source intensity parameter $\mu$ (see~\eqref{eq:concentration_diffusion}). 

\begin{remark}
	Consider the diffusion equation~\eqref{eq:concentration_diffusion} and a stationary arm with the range $\rho^\star (s), ~ s\in [0, L]$. As $t \rightarrow \infty$, the concentration $c(s,t)$ approaches the steady-state solution $c^\star (s) := -\tfrac{1}{\mu} \log \rho^\star (s)$. Using this equation, an estimate of the target distance is obtained as $\hat{\uprho}_i = e^{-\hat{\upmu}_i \c_i}$ (see Figure~\ref{fig:sensing_model}(b)).
\end{remark}

\vspace*{-10pt}
\paragraph{Mathematical definition of the sensing problem.} 
The sensing problem is to design the inputs $\gamma^{\uptheta_i}(t)$, $\gamma^{\upalpha_i}(t)$ for the two neural rings, and an update rule for $\hat{\upmu}_i (t)$ such that the estimates $\hat{\uptheta}_i (t)$, $\hat{\upalpha}_i (t)$, and $\hat{\upmu}_i (t)$ converge to their true values as $t \rightarrow \infty$.

\begin{table}[!t]
	\footnotesize
	\centering
	\caption{Nomenclature -- sensing model}
	\vspace{-8pt}
	\begin{tabular}{ll | ll | ll}
		\rowcolor{black}
		\multicolumn{2}{c}{\color{white} Sensory system} &
		\multicolumn{4}{c}{\color{white} Neural ring}\\
		$\alpha$ & target bearing & $\mathsf{V}$ & membrane potential & $w$ & weight function \\
		$\dist$ & target distance & $\tauRing$ & time constant & $\gamma$ & shifting factor (input) \\
		\rowcolor{black}
		\multicolumn{6}{c}{\color{white} i-th sensing unit} \\
		$s_i$ & arc-length location & $\mathcal{N}_i$ & set of adjacent suckers & $\c_i$ & chemical concentration input \\
		$\rsucker_i$ & position in lab frame & $\uptheta_i$ & orientation in lab frame & $\upkappa_i$ & curvature input \\
		$\uprho_i$ & true target distance & $\upalpha_i$ & true bearing & $\rt_i$ & estimated target location\\
		$\hat{\upalpha}_i$ & estimated bearing & $\hat{\upmu}_i$ & estimated intensity parameter & $\hat{\uptheta}_i$ & estimated orientation \\
		\hline
	\end{tabular}
	\label{tab:nomenclature_sensing}
\end{table}

\subsection{Sensing algorithm and its analysis} \label{sec:algorithm_sensing}

The $i$-th sensing unit maintains the estimates $(\hat{\uptheta}_i (t), \hat{\upalpha}_i (t), \hat{\upmu}_i (t))$. In terms of these variables, the estimated target location by the $i$-th sensing unit (in the lab frame) is defined as (see Figure~\ref{fig:sensing_model}(b))
\begin{align}
	{\rt}_i = \bm{\mathsf{r}}_i + e^{-\hat{\upmu}_i \c_i} \begin{bmatrix} \cos (\hat{\uptheta}_i + \hat{\upalpha}_i) \\ \sin (\hat{\uptheta}_i + \hat{\upalpha}_i)
	\end{bmatrix}, ~~ i = 1, 2,..., N
	\label{eq:target_vector_estimate}
\end{align} 
The collection of estimates are denoted as $\hat{\bm{\uptheta}} (t) = (\hat{\uptheta}_1 (t),  ..., \hat{\uptheta}_N (t))$,  $\hat{\bm{\upalpha}} (t) = (\hat{\upalpha}_1 (t),  ..., \hat{\upalpha}_N (t))$, and $\hat{\bm{\upmu}} (t) = (\hat{\upmu}_1 (t),  ..., \hat{\upmu}_N (t))$.

The proposed solution to the sensing problem is based on definition of two types of energy, $\mathcal{E}^{\text{prop}} (\cdot)$ and $\mathcal{E}^{\text{chemo}} (\cdot)$, for the two types of sensing modalities. 

\medskip
\noindent
\textit{Energy for proprioception} is defined as 
\begin{align}
	\mathcal{E}^{\text{prop}} = \mathcal{E}^{\text{prop}} (\hat{\bm{\uptheta}})  := \frac{k_\uptheta}{2}\sum\limits_{i = 1}^N \left(1 - \cos ( \hat{\uptheta}_i - \hat{\uptheta}_{i-1} - \upkappa_i \Delta s_{i, i-1} )\right)
	\label{eq:consensus_cost_proprioception}
\end{align}
where $k_\uptheta >0$ is a positive constant, $\hat{\uptheta}_0 = 0$ and $\Delta s_{1,0} = 0$. This energy penalizes $\hat{\uptheta}_i$ to deviate from $\hat{\uptheta}_{i-1} + \upkappa_i \Delta s_{i,i-1}$ which is a discrete approximation of the arm kinematics~\eqref{eq:kinematics}.
The notation $\nabla_{\hat{\uptheta}_i} \mathcal{E}^{\text{prop}}$ is used to denote the partial derivative of the right hand side with respect to $\hat{\uptheta}_i$.

\medskip
\noindent
\textit{Energy for chemosensing} is defined as
\begin{align}
	\mathcal{E}^{\text{chemo}} = \mathcal{E}^{\text{chemo}} (\hat{\bm{\uptheta}}, \hat{\bm{\upalpha}}, \hat{\bm{\upmu}}) := \frac{1}{2}\sum\limits_{i = 1}^N \sum\limits_{j \in \mathcal{N}_i} k_{\rsucker} \abs{ {\rt}_i - {\rt}_j}^2 + k_\upmu (\hat{\upmu}_i - \hat{\upmu}_j)^2
	\label{eq:consensus_cost_chemosensing}
\end{align}
where $k_{\rsucker}, k_{\upmu} >0$ are constant weight parameters.
The first term in \eqref{eq:consensus_cost_chemosensing} penalizes two neighboring sensors for predicting different target locations and the second term penalizes two neighboring suckers for predicting different values of the unknown parameter $\mu$. Additionally, $\nabla_{\hat{\uptheta}_i} \mathcal{E}^{\text{chemo}}$, $\nabla_{\hat{\upalpha}_i} \mathcal{E}^{\text{chemo}}$, $\nabla_{\hat{\upmu}_i} \mathcal{E}^{\text{chemo}}$ are used to denote the partial derivatives of the right hand side with respect to ${\hat{\uptheta}_i}, {\hat{\alpha}_i} , {\hat{\upmu}_i} $, respectively.

\medskip
We are now ready to state the estimation algorithm which is a type of consensus algorithm~\cite{olfati2007consensus, ren2008distributed} where the suckers communicate only with their neighbors in order to \textit{collectively} solve the sensing problem.

\vspace*{-10pt}
\paragraph{Consensus algorithm and its analysis.}
For the two neural rings with voltages $\mathsf{V}^{\uptheta_i} (\cdot, t)$ and $\mathsf{V}^{\upalpha_i} (\cdot, t)$, the respective inputs are obtained as follows:
\begin{subequations}
	\begin{align}
		\gamma^{\uptheta_i} (t) &= \tauRing \nabla_{\hat{\uptheta}_i} \mathcal{E}^{\text{prop}} \label{eq:theta_neural_ring_update}	 \\
		\gamma^{\upalpha_i} (t) &= \tauRing \left( \nabla_{\hat{\upalpha}_i} \mathcal{E}^{\text{chemo}} - \nabla_{\hat{\uptheta}_i}  \mathcal{E}^{\text{prop}} \right) \label{eq:alpha_neural_ring_update}	
	\end{align}
	and
	\begin{align}
		\frac{\dif \hat{\upmu}_i}{\dif t} &= - \nabla_{\hat{\upmu}_i} \mathcal{E}^{\text{chemo}}
		\label{eq:mu_update}
	\end{align}
	\label{eq:estimate_updates}
\end{subequations}
Explicit formulae of the update rules~\eqref{eq:estimate_updates} are provided in~\ref{appdx:sensing_algorithm}. Of particular interest are the terms involving the differences between position vectors of neighboring sensing units $(\rsucker_i - \rsucker_j), j \in \mathcal{N}_i, i = 1,..., N$. It is shown in~\ref{appdx:sensing_algorithm} that each of such terms is approximated by local shape angle estimates $\hat{\uptheta}_i$. In summary, the right-hand side is obtained entirely in terms of the estimated quantities, which in turn depend upon the sensory measurments.

The following theorem discusses the convergence properties of the consensus algorithm.

\begin{theorem}
	Suppose the arm is stationary with ground truth solutions for the two angles denoted by $\uptheta^\star_i$ and $\upalpha^\star_i$, $i = 1, 2, ..., N$. Suppose also that the sensory inputs are steady, denoted by $\upkappa_i (t) =: \upkappa^\star_i$ and  $\c_i (t) =: \c^\star_i$. Then,
	
	\smallskip
	\noindent
	A. {\bf (Proprioception)} The proprioceptory neural ring estimates asymptotically converge to
	\begin{align*}
		\hat{\uptheta}_i (t) \xrightarrow{(t \rightarrow \infty)}  \sum_{j=1}^i \upkappa^\star_j \Delta s_{j, j-1} =: \hat{\uptheta}_i^\star, ~~ i = 1, 2, ..., N
	\end{align*}
	
	(Note that the right hand side is a discrete approximation of the true angle $\uptheta_i^\star = \int_0^{s_i} \kappa^\star (s)\, \dif s$ obtained from kinematics equation~\eqref{eq:kinematics}.)
	
	\smallskip
	\noindent
	B. {\bf (Chemosensing)} Suppose further that $\hat{\upmu}_i=\mu$ for $i = 1, .., N$. Then the chemosensory neural ring estimates asymptotically converge to
	\begin{align*}
		\hat{\upalpha}_i (t) \xrightarrow{(t \rightarrow \infty)}  \upalpha_i^\star + (\uptheta_i^\star - \hat{\uptheta}_i^\star) =: \hat{\upalpha}_i^\star, ~~ i = 1, 2, ..., N
	\end{align*}
	\label{thm:consensus}
\end{theorem}

\begin{proof}
	See~\ref{appdx:consensus_proof}. 
\end{proof}

There are several implications of the chemosening consensus algorithm and the assumptions made in Theorem~\ref{thm:consensus}. These are discussed next in the form of following remarks.

\begin{remark} {\bf (Importance of proprioception)}
	The formula in Theorem~\ref{thm:consensus}-B is useful to see that the estimate of the shape angle is needed to estimate the bearing.
	This is true because the chemosensing energy $\mathcal{E}^{\text{chemo}}$ is a function of the sum of shape angle and bearing estimates $\hat{\uptheta}_i+\hat{\upalpha}_i$ (see~(\ref{eq:consensus_cost_chemosensing},\, \ref{eq:target_vector_estimate})).
	Physically, this signifies that the arm requires a sense of its own shape (internal sensation) in order to produce estimates of the local bearing (external sensation).
\end{remark}

\begin{remark} \label{remark:mu_estimate}
	{\bf (Role of the intensity parameter $\mu$)}	 
	In statement of Theorem~\ref{thm:consensus}-B, we make an assumption that $\hat{\upmu}_i = \mu$ for all $i$.
	These assumptions remove errors in range estimation, i.e. all the range estimates are exact ($\hat{\uprho}_i=\uprho_i$ for all $i$).
	An important goal of future work will be to analyze the effect of error in estimating $\mu$. Related problem appears in satellite navigation literature and referred to as dilution of precision (DOP)~\cite{langley1999dilution,spilker1996global,teunissen2017springer}. DOP means that the precision of localizing a given point is sensitive to the error in range measurements due to the relative geometry of the satellites (sensing units in our case). 
\end{remark}

Although the analysis of the consensus algorithm is carried out in a steady-state case, extensive numerical simulations are provided next (\S\,\ref{sec:sensing_results}) for the general time-dependent case.

\begin{figure*}[t]
	\centering
	\includegraphics[width=\textwidth, trim = {0pt 0pt 0pt 0pt}, clip = false]{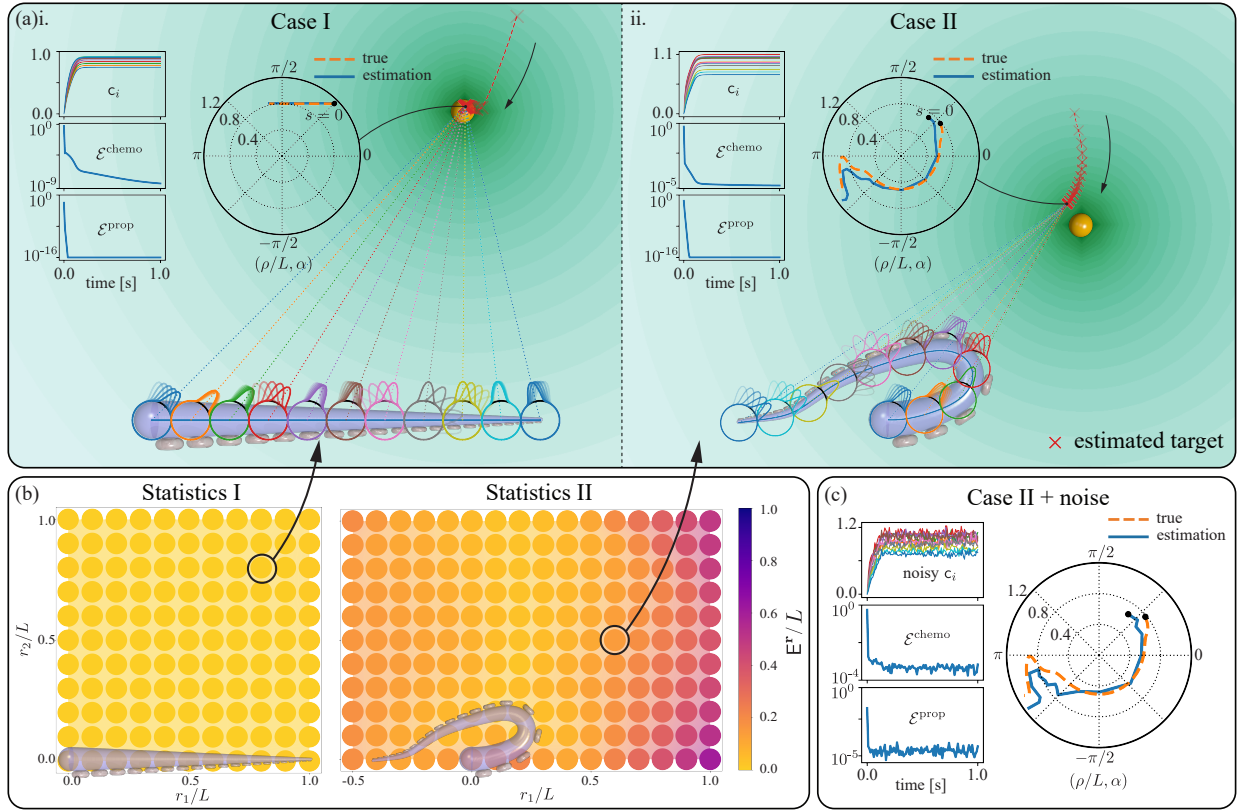}
	\vspace{-20pt}
	\caption{Sensing in octopus arm. (a)i. Case I: A straight arm senses a target located at $(0.8L,0.8L)$. Eleven neural rings are demonstrated among the total number of $N=21$ sensors. Each neural ring shows 5 frames of voltage peaks to demonstrate the shifts of angle $\hat{\uptheta}_i+\hat{\upalpha}_i$. The red cross with its trajectory illustrates the averaged location of the target estimates from all sensors. The first row of the inset on the left shows the trajectories of the concentration inputs of the 11 chosen sensors. The second and  third rows of the inset show the consensus cost over time for chemosensing and proprioception. The polar plot demonstrates the comparison between the estimation curve (blue solid line) and the true curve (orange dash line) of the pair $(\dist/L,\alpha)$. ii. Case II: A bent arm senses a target located at $(0.6L,0.5L)$. Same set of variables are illustrated as in Case I. (b) Statistics of sensing performance $\Er$ for a set of target locations with the straight arm (Statistics I) and the bent arm (Statistics II), respectively. (c) Case II with noisy inputs of concentration and curvature.}
	\label{fig:sensing_simulation}
	\vspace{-10pt}
\end{figure*}

\subsection{Simulation results for sensing} \label{sec:sensing_results}

This subsection presents the simulation results of the consensus algorithm proposed in \S\,\ref{sec:algorithm_sensing}. All parameter values are tabulated in Table~\ref{tab:parameters_sensing} and sensing results are presented in Figure~\ref{fig:sensing_simulation}. To assess the performance of the sensing algorithm, the following two cases are considered.

\smallskip
\noindent
\textbf{(1) Case I (straight arm):} In the first case, the arm is straight and the target is located at $\r^\target=(0.8L, 0.8L)$. Rapid convergence is seen with both types of consensus costs $\mathcal{E}^{\text{prop}}$ and $\mathcal{E}^{\text{chemo}}$ (see Figure~\ref{fig:sensing_simulation}(a)i). All estimates converge to their true values, i.e. $\hat{\uptheta}_i\rightarrow 0$, $\hat{\upalpha}_i\rightarrow\alpha(s_i)$, and $\hat{\upmu}_i\rightarrow \mu$ for $i=1,\dots,N$. A polar plot is used to compare the true curve and the estimation curve of $\Big(\dist(s)/L, \alpha(s)\Big)$ for $0\leq s\leq L$.  For Case I, the estimation curve overlaps the true curve.

\smallskip
\noindent
\textbf{(2) Case II (bent arm):} For the second case, the arm with an initial bend (defined by a curvature profile in~\cite[Sec. IV.A.3]{wang2021optimal}) is tested and the target is located at $\r^\target=(0.6L, 0.5L)$. The proprioception consensus cost approaches zero, indicating the convergence of the proprioception consensus algorithm. However, the chemosensing consensus cost does not converge to machine precision, indicating an error in chemosensing. The error is also seen in the polar plot in Figure~\ref{fig:sensing_simulation}(a)ii. This error is present because of multiple factors, including the estimation of parameter $\mu$ (see the discussion in Remark~\ref{remark:mu_estimate}).

For the two cases, the performance of the sensing algorithm is further investigated as a function of the target location and sensory noise. A total of 121 targets are selected on a uniform grid inside a square $[0,L]\times[0,L]$ for the straight arm (Statistics I), while a total of 176 targets are selected on a uniform grid inside a rectangle $[-0.5L,L]\times[0,L]$ for the bent arm (Statistics II). For each target, the following error metric is introduced
\begin{equation}
	\Er := \frac{1}{N} \sum_{i=1}^N\left|\r^\target - {\rt}_i\right|
\end{equation}
where ${\rt}_i$ is defined in~\eqref{eq:target_vector_estimate}. The error statistics for both cases are illustrated in Figure~\ref{fig:sensing_simulation}(b) and are tabulated in Table~\ref{tab:sensing_performance}. 
It is observed that the sensing algorithm achieves high level of accuracy for Case I (straight arm). However, for Case II (bent arm), the error in locating the target varies depending upon the target location. Specifically, high degrees of error occur for the targets located at the top-right and bottom-right corner of the rectangle.
These errors prevail because of the absence of simplifying assumptions in Theorem~\ref{thm:consensus}, including the estimation of $\mu$. It is shown in~\ref{appdx:statistics} that reinstating such assumptions improves the error statistics, as is prescribed by Theorem~\ref{thm:consensus}.

\begin{table}[!t]
	\footnotesize
	\centering
	\caption{Parameters for sensing and initialization}
	\hspace*{-5pt}
	\newcolumntype{b}{>{\columncolor{black}}c}
	\begin{tabular}{ccc|bbb}
		\rowcolor{black}
		{\color{white} Parameter} & 
		{\color{white} Description} & 
		{\color{white} Value} & 
		{\color{white} Variable} & 
		\multicolumn{2}{c}{\color{white}Initialization} \\
		\rowcolor{white}
		&&&&& \\ [-9pt]
		\rowcolor{white}
		& {\bf Sensing system modeling} & &
		$\hat{\uptheta}_i$ & \multicolumn{2}{c}{$\sim\text{Unif}([-0.1\pi,0.1\pi]^{N})$} \\
		\rowcolor{white}
		&&&&& \\ [-9pt]
		\rowcolor{white}
		$N$ & number of sensing units & $21$  &
		$\hat{\upalpha}_i$ & \multicolumn{2}{c}{$\sim\text{Unif}([0,\pi]^{N})$} \\
		\rowcolor{white}
		&&&&& \\ [-9pt]
		\rowcolor{white}
		$\tauRing$ & neural ring time constant [s] & $0.01$  &
		$\hat{\upmu}_i$ & \multicolumn{2}{c}{$\sim\text{Unif}([0.5\mu, 1.5\mu]^{N})$} \\
		\cline{1-3}\noalign{\smallskip}
		\rowcolor{white}
		& {\bf Consensus algorithm} & &
		$\mathsf{V}^{\uptheta_i}(\varphi)$ & \multicolumn{2}{c}{$\mathsf{V}^{\text{desired}}(\varphi-\hat{\uptheta}_i)$} \\
		\rowcolor{white}
		$\mu$ & source intensity parameter & $2.0$ &
		$\mathsf{V}^{\upalpha_i}(\varphi)$ & \multicolumn{2}{c}{$\mathsf{V}^{\text{desired}}(\varphi-\hat{\upalpha}_i)$} \\
		$k_\uptheta$ & weight parameter for $\hat{\uptheta}_i$ & $5\times10^4$ &
		{\color{white} Variable} & 
		\multicolumn{2}{b}{\color{white}Noisy inputs} \\
		\rowcolor{white}
		$k_{\rsucker}$ & weight parameter for ${\rt}_i$ & $4\times10^4$  &
		curvature & \multicolumn{2}{c}{$\upkappa_i(1 + 0.05\ \text{Normal}(0, 1))$} \\
		\rowcolor{white}
		$k_\upmu$ & weight parameter for $\hat{\upmu}_i$ & $4\times10^4$ &
		concentration & \multicolumn{2}{c}{$\c_i(1 + 0.05\ \text{Normal}(0, 1))$} \\
		\hline \\ [-9pt]
		\multicolumn{6}{l}{\footnotesize{Note: the local shape angle estimate of the first sensing unit (at the base of the arm) is set to $\hat{\uptheta}_0=0$.}} \\
		\multicolumn{6}{l}{\footnotesize{$\text{Unif}([a,b]^N)$ represents $N$ samples distributed uniformly between $a$ and $b$. $\text{Normal}(0,1)$ represents }} \\
		\multicolumn{6}{l}{\footnotesize{the normal distribution with mean $0$ and variance $1$.}}
	\end{tabular}
	\label{tab:parameters_sensing}
\end{table}

Finally, the effect of noisy inputs is considered and illustrated in Figure~\ref{fig:sensing_simulation}(c). The setting of Case II is considered. The sensory inputs -- concentration $\c_i (t)$ and curvatures $\upkappa_i (t)$ are corrupted with noise as follows. At each time $t$, Gaussian noise signals of mean zero and variance of 5\% of their respective absolute values are added to mimic a real-world noisy scenario. It is observed from Figure~\ref{fig:sensing_simulation}(c) that the consensus algorithm is able to estimate the target location fairly well with an error of $\Er = 0.153L$ (compared to $\Er = 0.144L$ without noise).

\begin{table}[!t]
	\footnotesize
	\centering
	\caption{Sensing performance}
	\hspace*{-5pt}
	\begin{tabular}{cccc}
		\rowcolor{black}
		{\color{white} Metric} & 
		{\color{white} Case I} & 
		{\color{white} Case II} &
		{\color{white} Case II + noise} \\
		$\mathcal{E}^\text{prop}(t=1)$ & $10^{-16}$ & $10^{-16}$ & $10^{-5}$ \\
		$\mathcal{E}^\text{chemo}(t=1)$ & $10^{-9}$ & $10^{-5}$ & $10^{-4}$ \\
		$\Er/L$ & $5.89\times10^{-3}$ & $1.44\times10^{-1}$ & $1.53\times10^{-1}$ \\
		\hline\noalign{\smallskip}
		\multicolumn{2}{c}{\bf Statistics I} &
		\multicolumn{2}{c}{\bf Statistics II} \\
		$\min\Er/L$ & $4.45\times10^{-4}$ & $\min\Er/L$ & $2.38\times10^{-2}$ \\
		$\max\Er/L$ & $1.43\times10^{-2}$ & $\max\Er/L$ & $6.02\times10^{-1}$ \\
		$\text{mean}\,\Er/L$ & $3.86\times10^{-3}$ & $\text{mean}\,\Er/L$ & $1.59\times10^{-1}$ \\
		\hline
	\end{tabular}
	\label{tab:sensing_performance}
\end{table}

\begin{algorithm}[t]
	\caption{End-to-end sensorimotor control algorithm}
	\label{algm:sensorimotor_controll}
	\begin{algorithmic}[1]
		\Require  Concentration $\c_i$ and curvature $\upkappa_i$ for $i=1,\ldots,N$.
		\Ensure rod movement $\r(s,t)$ and $\theta(s,t)$
		\State Initialize the rod $\r(s,0)$ and $\theta(s,0)$;
		\State Initialize the cables $V^\muscle(s,0)$ and $\Vadapt^\muscle(s,0)$;
		\State Initialize the sensors $\mathsf{V}^{\uptheta_i}(s,0)$, $\mathsf{V}^{\upalpha_i}(s,0)$, and $\hat{\upmu}_i(0)$ for $i=1,\ldots,N$.
		\While{not reaching the target}
		\State Update $\mathsf{V}^{\uptheta_i}(s,t)$ according to~(\ref{eq:neural_ring},\,\ref{eq:consensus_cost_proprioception},\,\ref{eq:theta_neural_ring_update}) and obtain $\hat{\uptheta}_i(t)$ according to~\eqref{eq:sucker_estimates};
		\State Update $\mathsf{V}^{\upalpha_i}(s,t)$ according to~(\ref{eq:neural_ring},\,\ref{eq:consensus_cost_chemosensing},\,\ref{eq:alpha_neural_ring_update}) and obtain $\hat{\upalpha}_i(t)$ according to~\eqref{eq:sucker_estimates};
		\State Update $\hat{\upmu}_i$ according to~(\ref{eq:consensus_cost_chemosensing},\,\ref{eq:mu_update});
		\State Get $\hat{\alpha}(s,t)$ according to~\eqref{eq:continuous_alpha_estimate} and $\hat{\bar{s}}(t)$ according to~\eqref{eq:continuous_sbar_estimate};
		\State Get current input $I^\muscle(s,t)$ according to~\eqref{eq:sensory_feedback_control_law_approximate};
		\State Update $V^\muscle(s,t)$ and $\Vadapt^\muscle(s,t)$ according to~\eqref{eq:cable_eq_full};
		\State Get muscle couple $u^\muscle(s,t)$ according to~\eqref{eq:neuromuscular_mapping};
		\State Update $\r(s,t)$ and $\theta(s,t)$ according to~\eqref{eq:dynamics}.
		\EndWhile
	\end{algorithmic}
\end{algorithm}

\section{End-to-end sensorimotor control}  \label{sec:algorithm_combined}
In this section, the feedback control law of the neuromuscular arm system~\eqref{eq:sensory_feedback_control_law} is combined with the estimates of the sensory system~\eqref{eq:estimate_updates} to obtain an end-to-end sensorimotor control algorithm. The algorithm is described in~\S\,\ref{sec:sensorimotor_algorithm} and simulation results are provided in~\S\,\ref{sec:algorithm_combined_results}.

\subsection{Algorithm} \label{sec:sensorimotor_algorithm}
The neuromuscular control system receives discrete values of local bearing estimates $\hat{\upalpha}_i(t), \hat{\upmu}_i$. 
Based on these, continuous estimates $\hat{\alpha}(s,t), \hat{\rho} (s,t)$ are created using linear interpolation, i.e.,
\begin{align}
	\begin{rcases}
		\hat{\alpha}(s,t) = \frac{1}{\Delta s_{i+1,i}} \left( \hat{\upalpha}_i (t) (s_{i+1} - s)  + \hat{\upalpha}_{i+1} (t) (s - s_i) \right)\\
		\hat{\rho}(s,t) = \frac{1}{\Delta s_{i+1,i}} \left( \hat{\uprho}_i (t) (s_{i+1} - s)  + \hat{\uprho}_{i+1} (t) (s - s_i) \right)
	\end{rcases}
	~~ \text{for}~s\in [s_i,s_{i+1}], ~ i = 1,..., N-1
	\label{eq:continuous_alpha_estimate}
\end{align}
where $\hat{\uprho}_i(t) = e^{-\hat{\upmu}_i(t) \c_i(t)}, i = 1,2,..., N$.
An estimate of the arc-length of the point on the arm closest to target is then computed as
\begin{align}
	\hat{{s}}(t) = \argmin_{s\in [0,L]} ~~\hat{\rho} (s,t)
	\label{eq:continuous_sbar_estimate}
\end{align} 
The certainty equivalence sensory feedback control law is as follows:
\begin{align}
	I^\muscle (s,t) = f^\muscle\left(\hat{\alpha} (s,t), \hat{{s}} (t) \right), \quad \muscle \in \mathcal{M} 
	\label{eq:sensory_feedback_control_law_approximate}
\end{align}
where the definitions of the functions $f^\muscle(\cdot, \cdot)$ are given in \eqref{eq:sensory_feedback_control_law}.
A  pseudo code of the algorithm is provided in Algorithm~\ref{algm:sensorimotor_controll}. The efficacy of the sensorimotor control algorithm is demonstrated through numerical simulations in \S\,\ref{sec:algorithm_combined_results}.

\begin{figure*}[t]
	\centering
	\includegraphics[width=\textwidth, trim = {0pt 0pt 0pt 0pt}, clip = false]{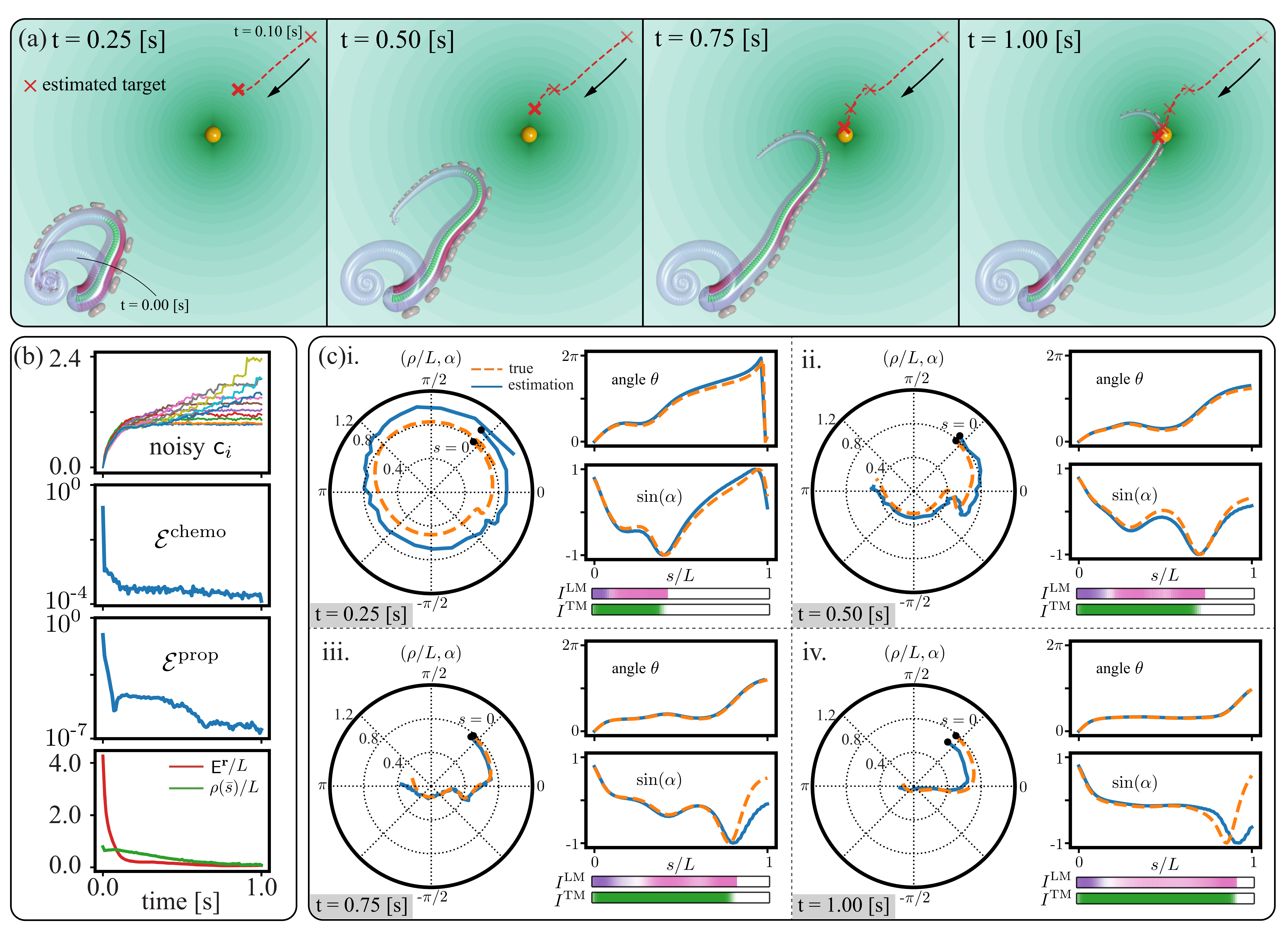}
	\vspace{-20pt}
	\caption{Sensorimotor control in octopus arm. (a) A bent arm with curl at the tip is initialized with a target presented at $(0.5L,0.6L)$. Five frames of moving arm demonstrates the reaching motion towards the target. The red cross shows the trajectory of the averaged target beliefs from all sensors. Eleven noisy sensory measurements of concentration over time are plotted in the first row of the inset on the left followed by two plots of consensus cost trajectories for chemosensing and proprioception. The inset on the bottom right shows the trajectories of $\Er$ (in red) for sensing performance and $\dist(\bar{s})/L$ (in green) for reaching performance. (b)i.-iv. Sensing and control statistics at 4 time instances, $t=0.25$ [s], $t=0.50$ [s], $t=0.75$ [s], and $t=1.00$ [s]. At each time instance, the following are compared between estimation (blue solid line) and true value (orange dash line): the polar plot for $(\dist/L,\alpha)$ curves, the sine of bearing $\sin(\alpha(s))$, and the arm local orientation $\theta(s)$. The current control inputs are illustrated in color bars with $I^{\LM}$ in purple and pink for top and bottom longitudinal muscles, respectively, and with $I^{\TM}$ in green. The transparency of the color bar indicates the magnitude of the control.}
	\label{fig:reaching_simulation}
	\vspace{-10pt}
\end{figure*}

\subsection{Simulation results for end-to-end sensorimotor control} \label{sec:algorithm_combined_results}

The arm is initialized at the same configuration described in \S\,\ref{sec:neuromuscular_results}. A static target at coordinates $(0.5L,0.6L)$ is presented. Sensory input noise is added as described in \S\,\ref{sec:sensing_results}. 
The parameters for the sensory system and control are chosen to be the same as those in Table~\ref{tab:parameters_sensing} and Table~\ref{tab:parameters_neuromuscular}, respectively.

Simulation results are portrayed in Figure~\ref{fig:reaching_simulation}. Four snapshots of the arm dynamic motion are illustrated in part (a) of the figure. The arm is observed to perform bend propagation toward the target, while estimating the target location simultaneously. Figure~\ref{fig:reaching_simulation}(b) shows the performance of the sensorimotor control where the effect of sensory noise is discernible. 
To further elucidate the dynamic evolution of the arm sensory and motor control systems, more details are shown for four time instances in Figure~\ref{fig:reaching_simulation}(c)i-iv. It is observed from the polar plots that the estimation curve of $(\dist/L,\alpha)$ tracks the true curve over time. The arm shape angle $\theta(s)$ and the $\sin(\alpha(s))$ curves also show that the estimates match the true values reasonably well, indicating efficacy of the sensing scheme. On the other hand, the control color bars over four time instances indicate a wave of muscle actuation which is also reflected in the bend propagation movement of the arm in Figure~\ref{fig:reaching_simulation}(a). Overall, with the sensing and the control updating simultaneously, the arm successfully reaches the target at $t=1.00$ [s].

\section{Conclusion and future work} \label{sec:conclusion}
This article presents neurophysiologically inspired models and algorithms for sensorimotor control of an octopus arm. The sensorimotor control problem for the arm is to catch a target that is emitting chemical signals. The arm has the ability to sense the chemical signal through its suckers (chemosensing) and its own curvature though its intramuscular proprioceptors (proprioception). 

First, a mathematical model for the octopus peripheral nervous system is proposed and assimilated with a nonlinear elastic model of the muscular soft arm.  Next, a neural-ring circuit model is proposed for generating sensory representations of different sensing modalities -- chemosensing (external) and proprioception (internal). Building on these mathematical models, several biophysically inspired algorithms are then proposed to solve the sensorimotor control problem: (i) a novel neuromuscular control law is shown to stabilize the arm while accomplishing the task of reaching to a given target; (ii) the sensing tasks of locating a target and estimating the shape of the arm are solved by a consensus algorithm in which the arm sensors communicate with their neighbors; and (iii) a combined sensorimotor control algorithm is shown to solve the sensing and control tasks simultaneously. Systematic numerical simulations demonstrate and support our theoretical analyses. Simulation results also reveal qualitative matches with behavioral observations of octopus arms, including curled rest shapes and target-oriented bend propagation movement. Further methodical analyses embellish the efficiency and robustness of our proposed algorithms. 

Our models, incorporating biomechanical and neural dynamics, not only reproduce observed behaviors but also provide a platform for hypothesis testing and refinement. They serve as valuable tools for deepening our understanding of the neural basis of octopus arm sensing and control. An important next step is to extend our framework to encompass three-dimensional arm movements~\cite{kennedy2020octopus,chang2023energy,tekinalp2023topology}. In a 3D environment, there is need for feedback control strategies for automatic twisting of the arm so as to align the suckers for maximum chemical signal reception. The sucker alignment control strategy could be used for object manipulation such as automatic grasping~\cite{tekinalp2023topology}. Another important next step would be resolving the flow induced by the arm motion and its feedback on the advection and diffusion of sensed chemicals~\cite{tekinalp2024soft}. On the other hand, mechanosensing is a significant sensing modality that the arm relies on for locomotion or object manipulation~\cite{levy2017motor,chang2023mechanosensory}. However, neural basis for mechanosensing and associated control strategies remain largely unknown~\cite{rowell1966activity, chang2023mechanosensory}. Our future work will attempt to reveal the mysteries of these intricate neural control mechanisms.

The models and algorithms developed in this article may potentially transcend the octopus arms by being applied in soft robotics~\cite{laschi2012soft, uppalapati2020berry}, where slender soft structures are equipped with actuators, such as artificial muscles~\cite{mirvakili2018artificial, xiao2019biomimetic} or pneumatic actuators~\cite{singh2017constrained, singh2020designing}; and sensors, such as cameras~\cite{albeladi2022hybrid, kamtikar2022visual}, magnetic sensors~\cite{bergamini2014estimating, kim2022physics}, and liquid metal sensors~\cite{xie2023octopus}. Sharing aspects pertinent to the octopus arms, efficient algorithms for local, distributed sensing and feedback control are of paramount importance for these systems.

\medskip
\noindent
\textbf{Acknowledgements.}~~{\footnotesize It is a pleasure to acknowledge Dr. P. S. Krishnaprasad of the University of Maryland, College Park for many insightful discussions on the topic of the octopus sensorimotor control problem. Authors are also thankful to Dr. William Gilly's lab at the Hopkins Marine Station, Stanford University, where the octopus reaching experiments were performed.}

\smallskip
\noindent
\textbf{Ethics.}~~{\footnotesize All octopus experiments were carried out in accordance with protocol \#23015 approved by the University of Illinois Urbana-Champaign (UIUC) Institutional Animal Care and Use Committee (IACUC).}

\smallskip
\noindent
\textbf{Authors’ Contributions.}~~{\footnotesize U.H., T.W., and P.G.M. contributed to the theoretical conceptualization of the problem. U.H and T.W. created the models of PNS, the sensory system, proposed the sensorimotor control algorithms, and carried out their theoretical analyses. T.W, U.H., and M.G. designed the simulations and interfacing with the package \textit{Elastica}. T.W. implemented the software. E.G. and R.G. performed the behavioral experiments, anesthesia, and histological analysis of an \textit{Octopus rubescens} arm. T.W. and E.G. rendered the figures. U.H. and T.W. wrote the manuscript. P.G.M., M.G., E.G., and R.G. helped edit the manuscript.}

\smallskip
\noindent
\textbf{Competing Interests.}~~{\footnotesize The authors declare no competing interests.}

\smallskip
\noindent
\textbf{Funding.}~~{\footnotesize The authors gratefully acknowledge financial support from ONR MURI N00014-19-1-2373, NSF EFRI C3 SoRo \#1830881, NSF OAC \#2209322, and ONR N00014-22-1-2569. The authors also acknowledge computing resources provided by the Extreme Science and Engineering Discovery Environment (XSEDE), which is supported by National Science Foundation grant number ACI-1548562, through allocation TGMCB190004.}

\appendix
\renewcommand\thesection{Appendix \Alph{section}} 
\renewcommand{\thesubsection}{\Alph{section}.\arabic{subsection}}
\renewcommand{\theequation}{\Alph{section}.\arabic{equation}}
\renewcommand{\thelemma}{\Alph{section}.\arabic{lemma}}
\setcounter{table}{0}
\renewcommand{\thetable}{\Alph{section}.\arabic{table}}

\section{Experimental protocols} \label{appdx:experiments}

\paragraph{Tissue Sectioning.} A selected part of isolated \textit{Octopus rubescens} arm was cut ($\sim$2 cm) and put in to 330 mM $\text{MgCl}_2$ solution for 30 minutes, allowing for muscle relaxation. The arm piece was then fixed in 3\% paraformaldehyde (PFA) in phosphate-buffered saline (PBS), $>10$X volume, overnight (for $\sim$ 8-9 hours) at $4^\circ$C, and transferred to PBS for long term storage at 4$^\circ$C. Octopus arm samples were sectioned using a Leica CM1900 cryostat, with cryochamber temperature set at $-21^\circ$C, and specimen head temperature at $-10^\circ$C.  Each sample was attached, in the desired orientation, to a frozen cryostat chuck using sectioning media (Neg-50™ Embedding Medium Blue) and left to freeze in the cryochamber for 15 minutes. Sections were cut in longitudinal and transverse orientations, at 60 µm thickness, floated in PBS, and later stored at~4$^\circ$C.

\vspace*{-10pt}
\paragraph{Behavioral experiments for sensing.} In Figure~\ref{fig:behavior}(b), the behavior of the suckers of a freely moving octopus arm is demonstrated when presented with a piece of shrimp. The suckers first orient toward the food source, followed by the arm’s movement to capture the food. Subsequently, magnesium chloride injections were used as a local anesthetic to reversibly disrupt communication between the central brain and the arm. This anesthetization procedure was utilized to test sucker and arm orienting responses to odor plumes (Figure~\ref{fig:behavior}(c)), revealing appetitive behavior similar to that observed in intact non-anesthetized
arms. Furthermore, as previously reported by Wang et al.~\cite{wang2022sensory}, sucker responses to chemical stimuli persist in
isolated arms. These behavioral experiments suggest that the suckers can determine the direction to the food source (bearing), with the underlying mechanisms yet to be fully understood. The details of the anesthetization procedure is described next.

\vspace*{-10pt}
\paragraph{Anesthesia.} \textit{Octopus rubescens} arm anesthetization was done by injecting the base of an octopus arm with 330 mM $\text{MgCl}_2$ solution with 10 mM HEPES, pH = 7.6~\cite{butler2018vivo}. This temporarily disrupts communication between the arm and octopus’s central nervous system (CNS). For arm isolation, animals were anesthetized in 1-2\% ethanol in sea-water~\cite{butler2018vivo}, an arm was anesthetized with $\text{MgCl}_2$, as above, and the anesthetized arm was cleanly isolated with a single-edge razor blade. The animals were closely monitored in their recovery.

\begin{table}[!t]
	\footnotesize
	\centering
	\caption{Parameters for numeircal simulation}
	\vspace{-8pt}
	\begin{tabular}{cccc}
		\rowcolor{black}
		{\color{white} Equations} & 
		{\color{white} Time integration method} & 
		{\color{white} Time step size} & 
		{\color{white} Spacing} \\
		\hline\noalign{\smallskip}
		diffusion equation~\eqref{eq:concentration_diffusion} & Euler method & $\Delta t=1\times10^{-5}$ [s] & $\Delta x=\Delta y=0.05L$ \\
		Cosserat rod dynamics~\eqref{eq:dynamics} & Position verlet & $\Delta t=1\times10^{-5}$ [s] & $\Delta s=0.01L$ \\
		cable equations~\eqref{eq:cable_eq_full} & Runge-Kutta method & $\Delta t=1\times10^{-5}$ [s] & $\Delta s=0.01L$ \\
		neural ring dynamics~\eqref{eq:neural_ring} & Euler method & $\Delta t=1\times10^{-5}$ [s] & $\Delta \varphi=0.02\pi$ \\
		\hline
		\multicolumn{4}{l}{\footnotesize{Note: finite difference approximation method is used for all spatial discretization.}}
	\end{tabular}
	\label{tab:numerics}
\end{table}

\section{Simulation environment setup} \label{appdx:numerics_setup}

This section describes the simulation environment setup for the following four different sets of equations: diffusion equation, Cosserat rod dynamics, cable equations, and neural rings. Details of integration schemes used for temporal discretization and spatial discretization are listed in Table~\ref{tab:numerics}.

\vspace*{-10pt}
\paragraph{Chemical signals from a source.}
To create the simulation environment, a point food-source is first placed at $\mathbf{r}^\target$ in a two-dimensional arena.
The emitted chemical concentration from the source diffuses through the water and is intercepted by the chemoreceptors of the arm suckers. The chemical diffusion is modeled as the diffusion equation presented in~\eqref{eq:concentration_diffusion} (see also discussions in Remark~\ref{remark:turbulence}).

The diffusion equation~\eqref{eq:concentration_diffusion} is numerically solved using Euler method with time step size $\Delta t=1\times10^{-5}$ [s] for time integration. Finite difference approximation is used with grid spacing $\Delta x=\Delta y=0.05L$ for spatial discretization.

\begin{remark} \label{remark:turbulence}
	The flow for odor plumes is known to be turbulent if the spatial scale is large enough\cite{baker2018algorithms, celani2014odor, murlis1992odor}. In fact, the turbulent nature of the odor plumes is a significant source of challenge in studying chemotactical behavior across species\cite{baker2018algorithms}. Since the turbulent flow are known to be analytically intractable, we restrict ourselves to a diffusive environment in this paper. This assumption may be justified by the proximity of the odor source and arm suckers in practice. Employing simplifying assumptions\cite{riman2021dynamics, hengenius2021olfactory}, we will study the turbulent nature of odor plumes and its effect in octopus olfaction in a future work. 
\end{remark}

\vspace*{-10pt}
\paragraph{Instantiating a soft muscular arm.}
The Cosserat rod dynamics~\eqref{eq:dynamics} are solved numerically by using the open-source software \textit{Elastica}~\cite{gazzola2018forward,zhang2019modeling,naughton2021elastica}. The temporal integration scheme uses the position verlet with time step size $\Delta t=1\times10^{-5}$ [s]. The continumn rod is discretized into $\Nelem$ connected cylindrical segments. The radius profile of a tapered rod is given by
\begin{equation}
	\radius(s) = \frac{s}{L}\rodtip + \frac{L-s}{L}\rodbase
	\label{eq:radius}
\end{equation}
which is based on measurements in real octopuses~\cite{chang2020energy}. The values of $\rodtip$ and $\rodbase$ are given in Table~\ref{tab:parameters_neuromuscular}. We compute the cross sectional area $A$ and second moment of area $\Isma$ as $A(s)=\pi(\radius(s))^2$ and $\Isma(s)=\frac{A(s)^2}{4\pi}$, respectively.
The rest of parameter values for the biophysically realistic arm are listed in Table~\ref{tab:parameters_neuromuscular}. A summary of notation and parameter values of muscle model is given in Table~\ref{tab:muscle_model}. Further details can be found in~\cite{gazzola2018forward,chang2021controlling}.

\vspace*{-10pt}
\paragraph{Simulating cable equations~\eqref{eq:cable_eq_full} for control.}
The neural dynamics~\eqref{eq:cable_eq_full} are numerically solved consistently with the temporal and spatial discretization of \textit{Elastica} for suitable assimilation, i.e., the time step size is $\Delta t=1\times10^{-5}$ [s] and the number of discretized spatial cable elements $\Ncable=\Nelem=100$. The temporal integration scheme uses the Runge-Kutta method. The values of the time constants $\tau=0.04$ [s] and $\tauAdapt=0.4$ [s] are taken in the range identified in~\cite{ekeberg1993combined}. The neuronal output function is $g(\cdot)=\max(0,\cdot)$. The activation function in the neuromuscular coupling equation~\eqref{eq:neuromuscular_mapping} is chosen as~\cite{ekeberg1993combined}
\begin{equation}
	\sigma(V) = \frac{1}{2}\left(1 + \tanh\left(-\frac{1}{40}\arctanh(-0.98(V-40))\right) \right)
	\label{eq:neuromuscular_mapping-numeric}
\end{equation}
The rest of the parameter values used are found in Table~\ref{tab:parameters_neuromuscular}.

\vspace*{-10pt}
\paragraph{Simulating neural rings~\eqref{eq:neural_ring} for sensing}
The neural ring dynamics~(\ref{eq:neural_ring},\,\ref{eq:weight_function}) described in~\S\,\ref{sec:sensing_neuro} are numerically solved using Euler method with time step size $\Delta t=1\times10^{-5}$ [s], and finite difference approximation by discretizing into $\Nring=\Nelem=100$ elements. The explicit form of the synaptic response function $h(\cdot)$ is given by~\cite{zhang1996representation}
\begin{equation}
	h(\mathsf{V}) = 6.34\ln^{0.8}(1+e^{10(\mathsf{V}+0.5)})
	\label{eq:synaptic_response-numeric}
\end{equation}
The details of Mexican hat function $\weight$ are described in~\ref{appdx:weight} and an illustration is given in Figure~\ref{fig:sensing_model}(a).

\section{Details of neuromuscular arm system} \label{appdx:neuromuscular}

\subsection{Passive elasticity}\label{appdx:passive}
The flexible arm is considered to be a linearly elastic rod whose passive elasticity is given by
\begin{equation}
	\mathbf{n}^\text{e} = EA (\nu_1-1)\, \mathbf{a} + GA\nu_2 \, \mathbf{b} ,\quad m^\text{e} = EI\kappa
\end{equation}
where $E$ and $G$ are the material's Young's and shear moduli, respectively.

\subsection{Drag model}\label{appdx:drag}
The drag model is closely based upon~\cite{yekutieli2005dynamic, wang2022control}.  
We write the drag forces as
\begin{equation}
	\mathbf{f}^{\text{drag}} = -\frac{1}{2}\varrho_{\text{water}} \left(A\mytan \upxi\mytan v_1|v_1|\mathbf{a} + A\myper \upxi\myper v_2|v_2|\mathbf{b} \right)
\end{equation} 
where $\varrho_{\text{water}}$ is the density of water, $A\mytan(s)=2\pi \radius(s)$ is 
the surface area of a unit length segment, and $A\myper=2\radius(s)$ is 
the projected area of the unit length segment in the plane perpendicular 
to the normal direction. Here $\radius(s)$ denotes the radius profile~\eqref{eq:radius} of the rod.  
The coefficients $\upxi\mytan$ and 
$\upxi\myper$ denote the tangential and perpendicular drag 
coefficients whose values are reported in Table~\eqref{tab:parameters_neuromuscular}.
Finally, $v_1$ and $v_2$ are the components of the velocity $\mathbf{r}_t$ in the material frame, i.e., $\mathbf{r}_t = v_1\mathbf{a} + v_2\mathbf{b}$. 

\subsection{Details of the muscle model}\label{appdx:muscle}

A generic muscle is denoted as $\muscle\in{\LMt, \LMb, \TM}$. In the laboratory frame, its position is given by
\begin{equation}
	\r^\muscle := \r + \bm{x}^\muscle
	\label{eq:muscle_pos}
\end{equation}
where $\bm{x}^\muscle$ is the muscle's relative position with respect to the center line. Note that for longitudinal muscles $\LM\in\{\LMt,\LMb\}$, we have $\bm{x}^{\LM}=\pm x^{\LM}\mathbf{b}$ (positive for $\LMt$ and negative for $\LMb$). For transverse muscle, $\bm{x}^{\TM}=\bm{0}$.

The function $\musclecoeff^\muscle(s,t)$ in the formula of the active muscle force $\mathbf{n}^\muscle(s,t)$~\eqref{eq:muscle_force} is given by
\begin{equation}
	\musclecoeff^\muscle(s,t) = \upsigma_{\max}^\muscle A^\muscle f_l(\ell^\muscle)
	\label{eq:hills_model}
\end{equation}
where $\upsigma_{\max}^\muscle$ is the maximum muscle force per unit area, $A^\muscle$ is the muscle cross sectional area, and function $f_l(\cdot)$ is a force-length relationship which takes argument $\ell^\muscle$, the local muscle length, that depends upon arm deformations given as follows (see \cite[Sec. III-D]{chang2021controlling} for details):
\begin{equation}
	\begin{aligned}
			\ell^{\LM} & = \r^{\LM}_s\mathbf{a} = \nu_1 \mp x^{\LM}\kappa \\
			\ell^{\TM} &= \frac{1}{\nu_1} \approx 2 - \nu_1
		\end{aligned}
\end{equation}
The force length curve $f_l(\cdot)$ is given by~\cite{chang2023energy}
\begin{equation}
	f_l(\ell) = \max\{3.06\ell^3 - 13.64\ell^2 + 18.01\ell - 6.44, 0\}
	\label{eq:f-l-relation-numeric}
\end{equation}
The rest of the notations of the muscle model and the values of parameters are listed in Table~\ref{tab:muscle_model}.

	\colorlet{shadecolor}{gray!40}
	\begin{table*}[!t] 
		\caption{Summarized muscle model}
		\hskip-10pt
		\centering
		\begin{tabular}{c ;{3.14159pt/2.71828pt} ccc ;{3.14159pt/2.71828pt} ccc} 
					\rowcolor{black}
					\hline
					& \multicolumn{3}{c}{\color{white} Muscle model notation} &  \multicolumn{3}{c}{\color{white} Values of parameters}\\
					\rowcolor{shadecolor}
					&&&&&&\\[-0.8em]
					\rowcolor{shadecolor}
					muscle & relative position & tangent vector & local length &  & max stress & normalized area  \\
					\rowcolor{shadecolor}
					$\muscle$ & $\bm{x}^\muscle$ & $\mathbf{t}^\muscle$ & $\ell^\muscle$ & \multirow{-2}{*}{$\dfrac{x^\muscle}{\radius}$} & $\upsigma^\muscle_{\max}$  [kPa] & $A^\muscle/A$ \\
					&&&&&&\\[-0.8em]
					$\LMt$ & $ x^\LM \, \mathbf{b}$ & $\mathbf{a}$ & $\nu_1-x^\LM\kappa$ & $\frac{5}{8}$ & 10 & $\frac{1}{8}$ \\
					&&&&&&\\[-0.8em]
					$\LMb$ & $- x^\LM \, \mathbf{b}$ & $\mathbf{a}$ & $\nu_1+x^\LM\kappa$ & $\frac{5}{8}$ & 10 & $\frac{1}{8}$ \\
					&&&&&&\\[-0.8em]
					$\TM$ & 0 & $-\mathbf{a}$ & $2-\nu_1$ & 0 & 25 & $\frac{1}{4}$ \\
					\hline
				\end{tabular}
		\label{tab:muscle_model}
	\end{table*}		

\section{Rest state of the arm}\label{appdx:rest_state}
An equilibrium of the cable equation~\eqref{eq:cable_eq_full} under no control ($\Imuscle = 0$) is obtained as
\begin{align}
\lambda^2 \Vmuscle_{ss}  - \Vmuscle - b g (\Vmuscle) = 0, ~~ \Vmuscle(0) = \Vmuscle_0, \Vmuscle (L) = \Vmuscle_L
\label{eq:equilibrium_no_control}
\end{align}

\begin{lemma} 
	Consider the voltage equilibrium equation \eqref{eq:equilibrium_no_control}. Then,
	
	a) 	If $\Vmuscle_0 \Vmuscle_L \geq 0$, then 
	\begin{align}
			\Vmuscle (s) = k_1^\muscle e^{\tfrac{s}{\hat{\lambda}}} + k_2^\muscle e^{-\tfrac{s}{\hat{\lambda}}} := f(s; k_1^\muscle, k_2^\muscle, \hat{\lambda})
			\label{eq:equilibrium_V_case1}
		\end{align}
	where $k_1^\muscle, k_2^\muscle, \hat{\lambda}$ are constants which depend on $\Vmuscle_0, \Vmuscle_L, \lambda$ and $b$.
	
	\smallskip
	b) If $\Vmuscle_0 \Vmuscle_L < 0$, then
	\begin{align}
			\Vmuscle (s) = \begin{cases}
					f(s; p_1^\muscle, p_2^\muscle, \hat{\lambda}_1), \quad 0 \leq s \leq s_\circ^\muscle \\
					f(s; q_1^\muscle, q_2^\muscle, \hat{\lambda}_2), \quad s_\circ^\muscle \leq s \leq L
				\end{cases}
			\label{eq:equilibrium_V_case2}
	\end{align}
	where $p_1^\muscle, p_2^\muscle, q_1^\muscle, q_2^\muscle, \hat{\lambda}_1, \hat{\lambda}_2$ are constants which depend on $\Vmuscle_0, \Vmuscle_L, \lambda$ and $b$. Furthermore, the zero-crossing point $s_\circ^\muscle$ is found by solving the following nonlinear equation 
	\begin{align}
			f_s (s_\circ^\muscle; p_1^\muscle, p_2^\muscle, \hat{\lambda}_1) = f_s (s_\circ^\muscle; q_1^\muscle, q_2^\muscle, \hat{\lambda}_2))
			\label{eq:zero-crossing-point}
		\end{align}
	\label{lemma:equilibrium_no_control}
\end{lemma}
\begin{proof}
	$a)$ Assume the solution $\Vmuscle(s)$ switches the sign. Then, it crosses zero at least twice. Assume the first and second crosses are at $s=s_{\circ1},s_{\circ2}$ such that $0 < s_{\circ1} < s_{\circ2} < L$, i.e., $\Vmuscle(s_{\circ1})=\Vmuscle(s_{\circ2})=0$. For $s\in(0,s_{\circ1})$, the solution $\Vmuscle(s)>0$ and is given by
	\begin{equation*}
			\Vmuscle(s) = \frac{\Vmuscle_0}{e^{\frac{s_{\circ1}}{\lambdaA}}-e^{\frac{-s_{\circ1}}{\lambdaA}}}\left(-e^{\frac{1}{\lambdaA}(s-s_{\circ1})} + e^{\frac{-1}{\lambdaA}(s-s_{\circ1})}\right)
		\end{equation*}
	where $\lambdaA$ takes value depending on the sign of the boundary values. For $s\in (s_{\circ1}, s_{\circ2})$, $\Vmuscle(s)\equiv0$ by solving the ODE~\eqref{eq:equilibrium_no_control} with zero Dirichlet boundary condition. However, the spatial derivative from left of $s_{\circ1}$ is
	\begin{equation*}
			\Vmuscle_s(s_{\circ1}^{-})=\frac{-2\Vmuscle_0}{\lambdaA\left(e^{\frac{s_{\circ1}}{\lambdaA}}-e^{\frac{-s_{\circ1}}{\lambdaA}}\right)} \neq 0
		\end{equation*}
	Thus, the solution $\Vmuscle(s)$ is not smooth at $s_{\circ1}$ and we reach a contradiction. Hence, the solution does not switch the sign. And the constants in the solution~\eqref{eq:equilibrium_V_case1} are given by
	\begin{equation*}
			k_1^\muscle = \frac{\Vmuscle_L - e^{\frac{-L}{\lambdaA}}\Vmuscle_0}{e^{\frac{L}{\lambdaA}}-e^{\frac{-L}{\lambdaA}}},\quad
			k_2^\muscle = \frac{e^{\frac{L}{\lambdaA}}\Vmuscle_0 - \Vmuscle_L}{e^{\frac{L}{\lambdaA}}-e^{\frac{-L}{\lambdaA}}},\quad \lambdaA=\left\{\begin{aligned}
					\lambda,\quad& \Vmuscle_0\leq0 \text{ and }\Vmuscle_L\leq0\\
					\frac{\lambda}{\sqrt{1+b}},\quad& \Vmuscle_0 \geq 0\text{ and }\Vmuscle_L \geq0
				\end{aligned}\right.
		\end{equation*}
	
	\smallskip
	$b)$ For $\Vmuscle_0\Vmuscle_L<0$, the solution $\Vmuscle(s)$ crosses zero at least once. Let us denote the first zero-crossing point as $s_\circ^\muscle$. In the interval $[s_\circ^\muscle, L]$, the problem boils down to the first case since we have $\Vmuscle\big|_{s=s_\circ^\muscle} \Vmuscle_L \geq 0$. It has been proved in part $a)$ that it cannot reach zero more than once. The constants in the solution~\eqref{eq:equilibrium_V_case2} are given by
	\begin{equation*}
			\begin{aligned}
					p_1^\muscle &= \frac{-e^{\frac{-s_\circ^\muscle}{\lambdaA_1}}\Vmuscle_0}{e^{\frac{s_\circ^\muscle}{\lambdaA_1}}-e^{\frac{-s_\circ^\muscle}{\lambdaA_1}}},\quad
					p_2^\muscle = \frac{e^{\frac{s_\circ^\muscle}{\lambdaA_1}}\Vmuscle_0}{e^{\frac{s_\circ^\muscle}{\lambdaA_1}}-e^{\frac{-s_\circ^\muscle}{\lambdaA_1}}},\quad \lambdaA_1=\left\{\begin{aligned}
							\lambda,&\quad \Vmuscle_0<0 \\ \frac{\lambda}{\sqrt{1+b}},&\quad \Vmuscle_0>0
						\end{aligned}\right. \\
					q_1^\muscle &= \frac{-e^{\frac{-s_\circ^\muscle}{\lambdaA_2}}\Vmuscle_L}{e^{\frac{s_\circ^\muscle-L}{\lambdaA_2}}-e^{\frac{L-s_\circ^\muscle}{\lambdaA_2}}},\quad
					q_2^\muscle = \frac{e^{\frac{s_\circ^\muscle}{\lambdaA_2}}\Vmuscle_L}{e^{\frac{s_\circ^\muscle-L}{\lambdaA_2}}-e^{\frac{L-s_\circ^\muscle}{\lambdaA_2}}},\quad \lambdaA_2=\left\{\begin{aligned}
							\lambda,&\quad \Vmuscle_L<0 \\ \frac{\lambda}{\sqrt{1+b}},&\quad \Vmuscle_L>0
						\end{aligned}\right.
				\end{aligned}
		\end{equation*}
	and the zero-crossing point $s_\circ^\muscle$ is found by solving~\eqref{eq:zero-crossing-point} with the above constants.
\end{proof}

\section{Motion camouflage strategy for a unicycle}\label{appdx:motion_camouflage}
As is pointed out in Remark~\ref{remark:motion_camouflage}, our proposed control law~\eqref{eq:sensory_feedback_control_law} is inspired by the models of the motion camouflage steering strategy~\cite{glendinning2004mathematics, justh2006steering}. A concise summary of the planar motion camouflage and its relationship to the octopus feedback control are discussed as follows.

\vspace*{-10pt}
\paragraph{Motion camouflage on a plane.}
Consider a point particle (pursuer) on a plane pursuing an evading target. Denote the position and orientation as $(\unicycle{x} (t), \unicycle{y}(t))$, and $\thetaUnicycle (t)$. Assume the pursuer moves at a constant speed $\unicycle{v}$ and the only control is the steering rate $\upomega$. The dynamics of the pursuer are described by the following unicycle system:
\begin{equation}
	\begin{aligned}
		\dot{\unicycle{x}} = \unicycle{v} \cos \thetaUnicycle, ~~\dot{\unicycle{y}} = \unicycle{v} \sin \thetaUnicycle, ~~
		\dot{\thetaUnicycle} = \upomega
	\end{aligned}
	\label{eq:dynamics_unicycle}
\end{equation}
Here, the dot notation is used for time derivatives. The moving target's dynamics can be represented in a similar way. We assume the target is moving at a constant speed $\unicycle{v}^\target$.

Let $\upzeta(t)$ be the distance between the pursuer and the target, $\upphi (t)$ be the bearing to the target with respect to the pursuer, and $\uppsi (t)$ be the bearing to the pursuer with respect to the target. Then the time evolution of $(\upzeta, \upphi)$ can be written as \cite{halder2016steering}
\begin{align}
	\begin{split}
		\dot{\upzeta} &= - \unicycle{v} \cos \upphi - \unicycle{v}^\target \cos \uppsi \\
		\dot{\upphi}     &= - \upomega + \frac{1}{\upzeta} \left( \unicycle{v} \sin \upphi + \unicycle{v}^\target \sin \uppsi \right) 
	\end{split}
	\label{eq:bearing_dynamics}
\end{align}

The motion camouflage control law \cite{justh2006steering} is the steering control given by
\begin{equation}
	\upomega = \upchi \left(\sin\upphi + \frac{\unicycle{v}^\target}{\unicycle{v}}\sin\uppsi  \right)
	\label{eq:ctrl_law_MC}
\end{equation}
where $\upchi>0$ is some large enough given constant.

\vspace*{-10pt}
\paragraph{Relationship to octopus arm feedback control.}
By interchanging the spatial variable for the arm with the temporal variable for the pursuit trajectory, parallels can be drawn between the arm equilibrium spatial configuration and the temporal trajectory of a pursuer intercepting a prey under the motion camouflage strategy. In particular, consider the case where the pursuer is moving at a unit speed ($\unicycle{v} \equiv 1$) towards a static target ($\unicycle{v}^\target \equiv 0$). In this case, from equations~\eqref{eq:bearing_dynamics} and \eqref{eq:sensory_kinematics_full} it is easy to see that the temporal variables $(\upzeta, \upphi)$ for a unicycle system are analogous to the spatial variables $(\rho, \alpha)$ for an inextensible and unshearable arm ($\nu_1 \equiv 1, \nu_2 \equiv 0$). Moreover, in this case, the motion camouflage control law~\eqref{eq:ctrl_law_MC} can be viewed as a parallel to our proposed feedback control laws~(\ref{eq:control_LM_top},\,\ref{eq:control_LM_bottom}).

\section{Proof of Theorem~\ref{thm:sensoryfeedback_control}} \label{appdx:sensoryfeedback_control_proof}

\subsection{Proof of Part A (Equilibrium analysis)} \label{appdx:sensoryfeedback_control_equilibrium_proof}

For the proof of this theorem, we ignore the superscript $^\star$ for the equilibrium variables for simplicity of notation.

With the assumption of inextensible and unshearable rod, transverse muscle is not relevant. Then, following the equilibrium analysis in \S\,\ref{sec:control_problem}, the equilibrium of the reduced system is given by the following set of equations:  

\noindent
(i) kinematics of the arm~\eqref{eq:kinematics},

\noindent
(ii) statics of the rod determined by only the boundary conditions~\eqref{eq:arm_boundary_conditions} and the curvature given as follows:
\begin{equation}
\kappa(s) =  \sum_{\muscle\in\{\LMt, \LMb\}} \frac{x_2^\muscle}{E\Isma}\musclecoeff^\muscle(s;\kappa)\sigma(V^\muscle(s))
\end{equation}

\noindent
(iii) the neuromuscular coupling $u^\muscle = \sigma (V^\muscle)$~\eqref{eq:neuromuscular_mapping},

\noindent
(iv) statics of the cables~\eqref{eq:cable_eq_full} with the free-free boundary condition~\eqref{eq:statics_cable}.

\noindent
(v) the reduced version of sensory kinematics~\eqref{eq:sensory_kinematics_full} with same boundary condition at $s=0$:
\begin{equation}
\begin{aligned}
		\dist_s(s) &= -\cos(\alpha (s)), \quad \rho(0) = \abs{\r^{\text{target}}}  \\
		\alpha_s(s) &= -\kappa(s) + \frac{1}{\dist (s)}\sin(\alpha (s)), \quad \mathsf{R}(\alpha (0))\, \mathbf{e}_1 = \frac{\r^{\text{target}}}{\rho(0)}. 
	\end{aligned}
\label{eq:sensory_kinematics}
\end{equation}
where $(\rho(0), \alpha(0))$ are calculated from the definition~\eqref{eq:definition-dist-bearing} and the arm's fixed boundary condition at $s=0$ as given in~\eqref{eq:arm_boundary_conditions}.

Notice that the curvature of the arm  $\kappa$ uniquely determines the equilibrium configuration $(\r, \theta)$ of the arm by integrating the kinematics~\eqref{eq:kinematics} with $\nu_1=1$ and $\nu_2=0$.

\smallskip
At the outset, define $\Gamma(s) :=\dist_s=-\cos\alpha$ and denote $\dist_0=\dist(0)$. 

For case (a), we necessarily have $\bar{s}\leq L$. The proof is completed in the following two steps:

\medskip
\noindent
\textit{Step 1:}
Define $\dist_1\in(0,\dist_0)$ such that $\chi=\frac{1}{\dist_1}+\const$ for some $\const>0$. Then, for $\forall \dist \in[\dist_1, \dist_0]$, we have
\begin{equation*}
\begin{aligned}
		\Gamma_s &= -\sin(\alpha)\kappa + \frac{1}{\dist}\sin^2(\alpha) \\
		&= -\sin^2(\alpha)\left(\chi-\frac{1}{\dist}\right) - \sin(\alpha)\left(\kappa-\chi\sin(\alpha)\right) \\
		&\leq -\const(1-\Gamma^2) + |\sin(\alpha)|\left|\kappa-\chi\sin(\alpha)\right|
	\end{aligned}
\end{equation*}

\noindent
\textbf{Claim: } $\left|\kappa-\chi\sin(\alpha)\right|$ is upper-bounded (denoted by $M$).

\begin{proof}
	\begin{equation*}
		\begin{aligned}
			\left|\kappa-\chi\sin(\alpha)\right| &\leq \left|\sum_{\muscle\in\{\LMt, \LMb\}} \frac{x_2^\muscle}{E\Isma}\musclecoeff^\muscle(s;\kappa)\right|\left|\sigma(V^\muscle)\right| + \left|I^{\LMt} - I^{\LMb}\right| \\
			&\leq \left(\max_{s\in[0,L]} \frac{2x^{\LM}}{E\Isma(s)}\musclecoeff^\muscle(s;\kappa(s))\right) \cdot 1 + \chi =:M
		\end{aligned}
	\end{equation*}
\end{proof}

\noindent
Then, for $1-\Gamma^2>\epsilon^2$, we have
\begin{equation*}
\begin{aligned}
		\Gamma_s &\leq -\const(1-\Gamma^2) + \frac{(1-\Gamma^2)}{\sqrt{1-\Gamma^2}}M \\
		&\leq -\left(\const - \frac{M}{\epsilon}\right)(1-\Gamma^2),\quad \forall \dist\geq\dist_1
	\end{aligned}
\end{equation*}
Note that $\dist_s=-\cos\alpha\geq-1$, which implies
\begin{equation*}
\begin{aligned}
		\dist(s)\geq\dist_1\,\ \forall s\leq \dist_0-\dist_1=:\Delta\rho_{01}
	\end{aligned}
\end{equation*}
Therefore, it is guaranteed that $\Gamma_s\leq0,\ \forall s\leq \Delta\rho_{01}$. By separation of variables and some calculations, we may derive
\begin{equation*}
\Gamma(s) \leq \tanh\left(\tanh^{-1}\Gamma_0 - \const s\right), ~~ \forall s \leq \Delta\rho_{01}
\end{equation*}
where we denote $\Gamma_0=\Gamma(0)$. We can therefore conclude that $\Gamma(\Delta\rho_{01})\leq\tanh\left(\tanh^{-1}\Gamma_0 - \const \Delta\rho_{01}\right)$. 

Note that for some $\epsilon_1 >0$ sufficiently small, $\tanh(z)\leq-1+\epsilon_1\Leftrightarrow z\leq\frac{1}{2}\ln\left(\frac{\epsilon_1}{2-\epsilon_1}\right)=:z_0$. Thus, if we take $\const$ to be sufficiently large such that
\begin{equation*}
\const \geq \frac{\tanh^{-1}\Gamma_0 - z_0}{\Delta\rho_{01}} + \frac{M}{\epsilon} =: \const_1
\end{equation*}
then we are guaranteed to achieve $\Gamma(\Delta\rho_{01})\leq-1+\epsilon_1$ given any small $\epsilon_1>0$ by using large enough $\chi=\frac{1}{\dist_1}+\const$.

\medskip
\noindent
\textit{Step 2:}
Note that
\begin{equation*}
\begin{aligned}
		\Gamma_s &= -\sin^2\alpha\left(\chi-\frac{1}{\dist}\right) 
		= -\sin^2\alpha\left(\frac{1}{\dist_1}+\const-\frac{1}{\dist}\right)\\
		&\leq 0, ~~ 
		\forall \ \dist \geq \frac{\dist_1}{\const\dist_1 + 1} =: \dist_2
	\end{aligned}
\end{equation*}
Similar to Step 1, we have $\dist(s)\geq\dist_2\,\ \forall s\leq \dist_0-\dist_2=:\Delta\rho_{02}$. For any $\epsilon>0$, choose $\dist_2\leq\epsilon$, i.e.,
\begin{equation*}
\const \geq \frac{1}{\epsilon} - \frac{1}{\dist_1} =: \const_2
\end{equation*}
Then, by taking $\const\geq \max\{\const_1, \const_2\}$, we have $\dist_s=\Gamma(s)\leq \Gamma(\Delta\rho_{01}) \leq -1+\epsilon_1,\ \forall \Delta\rho_{01}\leq s\leq \Delta\rho_{02}$. Then, for any $\epsilon>0$, we have
\begin{equation*}
\footnotesize
\begin{aligned}
		\dist(\Delta\rho_{02}) &= \dist_0 + \int_0^{\Delta\rho_{01}}\dist_s\ud s + \int_{\Delta\rho_{01}}^{\Delta\rho_{02}}\dist_s\ud s \\
		&\leq \dist_0 + \int_0^{\Delta\rho_{01}} \tanh\left(\tanh^{-1}\Gamma_0 - \const s\right) \ud s + (-1 + \epsilon_1)(\Delta\rho_{02} - \Delta\rho_{01}) \\
		&\leq \underbrace{\dist_0 + \frac{\ln\left(\frac{\cosh(b)}{\cosh(b-\const \Delta\rho_{01})}\right)}{\const}}_{\mathsf{h}(\const)} + (-1+\epsilon_1)\frac{\const\dist_1^2}{\const\dist_1+1} \leq \epsilon
	\end{aligned}
\normalsize	
\end{equation*}
where $b = \tanh^{-1} \Gamma_0$. One can derive that $\mathsf{h}'(\const)<0$ for $\const\geq \const_1$ and $\mathsf{h}(\const)\rightarrow \dist_1$ for $\const\rightarrow \infty$.
Then, $\exists \const_3 >0$ s.t. $\dist(\Delta\rho_{02})\leq\epsilon$ for $\const\geq \const_3$ and small enough $\epsilon_1$.

Note that for $s\in[\Delta\rho_{02},\bar{s}]$, we have $\chi-\frac{1}{\dist(s)}\leq 0$, $\alpha(s)\in(0,\frac{\pi}{2}]$ and thus, $\alpha_s>0$. Moreover, $\dist_s=-\cos\alpha\leq 0$ for $s\in[\Delta\rho_{02},\bar{s}]$. Hence, $\dist(\bar{s})\leq\dist(\Delta\rho_{02})\leq \epsilon$ by having $\const\geq \max\{\const_1,\const_2,\const_3\}$.

In conclusion, for all $\epsilon>0$, choose $\dist_1\in(0,\dist_0)$, $\exists \epsilon_1,\const$ s.t. $\epsilon_1>0$ and $\const=\max\{\const_1,\const_2,\const_3\}$. Then, there exits $\bar{\chi}=\bar{\chi}(\epsilon)=\frac{1}{\dist_1}+\const(\epsilon)>0$ such that for all $\chi>\bar{\chi}$, $\dist(\bar{s})\leq\epsilon$ for given $\bar{s}\leq L$.

\medskip
(ii) For this case, we have $\dist_0>L$ and $\bar{s}=L$. The proof is immediate by Step 1 in case (i) by choosing $\dist_1=\dist_0-L$, i.e. $\Delta\rho_{01} = \bar{s} = L$.

\subsection{Proof of Part B (Dynamic stability)} \label{appdx:sensoryfeedback_control_dynamics_proof}
We first separate the time-scales of the muscular arm dynamics~\eqref{eq:dynamics} and the neural dynamics~\eqref{eq:cable_eq_full} by noting that the neural system is `fast' because of the small time constants $\tau, \tilde{\tau}$ (also verified numerically). The `slow' mechanical system can be regarded as frozen with respect to the fast electrical system. Then, the Theorem~\ref{thm:sensoryfeedback_control}-B is proved by using the method of singular perturbation, as developed in~\cite[Chap. 11]{khalil2002nonlinear}. 

The proof is done in three steps. For simplicity, let us consider an inextensible and unshearable arm. Then, only the longitudinal muscles are involved which generates internal couple. Hence, we remove the superscripts $\muscle$ for notational ease. Let $\pr=\varrho A\r_t$ and $\ptheta=\varrho\Isma\theta_t$ be the translational and the angular momentum, respectively.

\begin{lemma}
Consider the rod dynamics~\eqref{eq:dynamics} with some couple control, $u=u(\kappa)$, as a given function of the curvature profile $\kappa$. Then the closed loop rod dynamics is (locally) asymptotically stable at the equilibrium $(\r,\theta, \pr, \ptheta)=(\r(\bar{\kappa}), \theta(\bar{\kappa}), \mathbf{0}, 0)$ defined by the curvature $\bar{\kappa}(s)=u(\bar{\kappa}(s))$.
\label{lemma:rod_stability}
\end{lemma}

\begin{proof}
	In \cite[Sec. III-E]{chang2021controlling}, it has been shown that if the internal muscle forces and couples are expressible as gradients of an  energy function (called \textit{muscle stored energy function}), then the rod system maintains its Hamiltonian structure (with damping) and (local) convergence to an equilibrium can be readily shown. In the present case, we see that the internal elastic couple is gradient of a quadratic elastic stored energy function $\storedenergy^{\text{e}} = \tfrac{1}{2} EI \kappa^2$. Assume the longitudinal muscle activation $u$ is a function of some given curvature profile $\kappa$, i.e. we may express $u=u(\kappa)$. Then define $\storedenergy^\muscle = \int u(\kappa)\ud \kappa$. Then it is clear that $u$ is the gradient of the function $\storedenergy^{\text{m}}$. Denote the kinetic energy, the elastic potential energy, and the muscle potential energy of the rod as $\kinetic$, $\potential^\text{e}$ and $\potential^\muscle$. Then, the total energy of the rod is given by the total controlled Hamiltonian as follows:
	\begin{equation}
		\Hamiltonian = \kinetic + \potential^\text{e} + \potential^\muscle
	\end{equation}
	We have
	\begin{equation}
		\frac{\ud \Hamiltonian}{\ud t} \leq -\xi\left\|\frac{1}{\varrho\Isma}\ptheta\right\|^2 \leq 0
	\end{equation}
	where the norm is taken in the $L^2$ sense. Thus, the total energy of the rod is non-increasing. The LaSalle's theorem guarantees the (local) asymptotic stability to the largest invariant subset of rod states with $\frac{\ud \Hamiltonian}{\ud t}=0$ which is indeed given by solving $\kappa=u(\kappa)$ for the curvature $\bar{\kappa}$.
\end{proof}

\begin{lemma}
Consider the neural dynamics~\eqref{eq:cable_eq_full} with some constant current control $I(s)$. Then the closed loop neural dynamics is (locally) asymptotically stable if $0<b\leq1$ and exponentially stable if $0<b<1$ at the equilibrium $(V,W)=(\Vref, \Wref)$ where $\lambda^2\Vref_{ss}-\Vref-\Wref+I=0$ and $\Wref=bg(\Vref)$.
\label{lemma:cable_stability}
\end{lemma}

\begin{proof}
	We propose the following Lyapunov candidate for the closed loop neural system:
	\begin{equation*}
		\small
		\LyapunovV (V, \Vadapt) = \frac{1}{2}\int_0^L\ \tau\left(V-\Vref\right)^2 + \tauAdapt\left(W-\Wref\right)^2 \ud s
	\end{equation*}
	It can be readily derived that
	\begin{equation}
		\scriptsize
		\begin{aligned}
			\frac{\ud \LyapunovV}{\ud t} &= \int_0^L\ \left(V-\Vref\right)\left(\lambda^2V_{ss}-V-W - \lambda^2\Vref_{ss}+\Vref+\Wref\right) + \left(W-\Wref\right)\left(-W+bg(V)\right) \ud s \\
			&\leq \int_0^L - \lambda^2\left(V_s-\Vref_s\right)^2 - \left(V-\Vref\right)^2 - \left(V-\Vref\right)\left(W-\Wref\right) \\
			&\qquad -\left(W-\Wref\right)^2 + \left(W-\Wref\right)\left(bg(V)-bg(\Vref)\right)\ud s \\
			&\leq - \lambda^2\left\|V_s-\Vref_s\right\|^2 - \left\|V-\Vref\right\|^2 + \left\|V-\Vref\right\|\left\|W-\Wref\right\| \\
			&\qquad -\left\|W-\Wref\right\|^2 + b\left\|W-\Wref\right\|\left\|g(V)-g(\Vref)\right\|\ud s \\
			&\leq - \lambda^2\left\|V_s-\Vref_s\right\|^2 - \left\|V-\Vref\right\|^2 + (1+b)\left\|V-\Vref\right\|\left\|W-\Wref\right\| -\left\|W-\Wref\right\|^2 \\
		\end{aligned}
	\end{equation}
	For $0<b\leq1$, we have $\LyapunovV\geq0$ and $\frac{\ud \LyapunovV}{\ud t}\leq0$ and the equal signs are taken only at the equilibrium $(V,W)=\left(\Vref, \Wref\right)$, hence (locally) asymptotically stable.
	
	Furthermore, for a constant $k=\frac{(\tau+\tauAdapt) - \sqrt{(\tau+\tauAdapt)^2-\tau\tauAdapt(4-(1+b)^2)}}{2\tau\tauAdapt}$, we have
	\begin{equation}
		\begin{aligned}
			\frac{\ud \LyapunovV}{\ud t} &\leq - \lambda^2\left\|V_s-\Vref_s\right\|^2 - k\tau \left\|V-\Vref\right\|^2 - k\tauAdapt \left\|W-\Wref\right\|^2 \\
			&\qquad -\left(\sqrt{1-k\tau}\left\|V-\Vref\right\| - \sqrt{1-k\tauAdapt}\left\|W-\Wref\right\|\right)^2 \\
			&\leq - \lambda^2\left\|V_s-\Vref_s\right\|^2 - k\tau \left\|V-\Vref\right\|^2 - k\tauAdapt \left\|W-\Wref\right\|^2 \\
			&\leq -2k\left(\frac{1}{2}\tau \left\|V-\Vref\right\|^2 + \frac{1}{2}\tauAdapt \left\|W-\Wref\right\|^2 \right) \\
			&\leq -2k\LyapunovV \nonumber
		\end{aligned}
	\end{equation}
	and thus the equilibrium is proved to be exponentially stable. Note that for $0<b<1$, $0<k<\frac{1}{\tau}$ is guaranteed. When $b=1$, we have $k=0$.
\end{proof}

\medskip
Finally, we come to the proof of Theorem~\ref{thm:sensoryfeedback_control}-B. Consider the coupled system~(\ref{eq:dynamics}, \ref{eq:cable_eq_full}, \ref{eq:neuromuscular_mapping}, \ref{eq:sensory_feedback_control_law}) for an inextensible and unshearable rod. In the coupled system, the current control input $I$ for the subsystem~\eqref{eq:cable_eq_full} is now a function of the bearing $\alpha$ which can be uniquely determined by some curvature profile $\kappa$ for any time $t$. Thus, we can write $I=I(\kappa)$. For any given curvature profile $\kappa$, denote the desired equilibrium of the single subsystem~\eqref{eq:cable_eq_full} as $(\Vref,\Wref)=(\Vref(\kappa), \Wref(\kappa))$. And further denote the desired couple $\uref=\sigma(\Vref)=\uref(\kappa)$. Note that $\LyapunovR$ is well-defined in the manifold $u=\uref(\kappa)$ from Lemma~\ref{lemma:rod_stability}.

For simplicity of writing, we define the following notations: $x=\left(\begin{smallmatrix}
	r \\ \theta \\ \pr \\ \ptheta
\end{smallmatrix}\right)$, $z=\left(\begin{smallmatrix}
	V \\ W
\end{smallmatrix}\right)$, and $y = z - \tilde{z} = z - h(x)$ where $h(x)=\tilde{z}=\left(\begin{smallmatrix}
	\Vref(\kappa) \\ \Wref(\kappa)
\end{smallmatrix}\right)$. Denote the dynamics~\eqref{eq:dynamics}\eqref{eq:neuromuscular_mapping} as $\frac{\ud x}{\ud t}=:f(x,z)$, and the dynamics~\eqref{eq:cable_eq_full}\eqref{eq:sensory_feedback_control_law} as $\varepsilon\frac{\ud z}{\ud t}=:g(x,z)$ where $\varepsilon=\left(\begin{smallmatrix}
	\tau \\ \tauAdapt
\end{smallmatrix}\right)$. Let $\psi_1(x):=\left\|\frac{1}{\varrho\Isma}\ptheta\right\|$, $\psi_2(y):=\left\|V_s-\Vref_s\right\|$, $\psi_3(y):=\left\|V-\Vref\right\|$, and $\psi_4(y):=\left\|W-\Wref\right\|$. Note that the external muscle couple is $\left(m(x,u)\right)_s=:\left(\musclecoeff(x)u\right)_s=\musclecoeff_su+\musclecoeff u_s$. Assume the following terms are bounded: $\left\|\frac{\partial h}{\partial x}\right\|\leq \beta_1$, $\left\|\musclecoeff_s\right\|\leq \beta_2$, and $\left\|\musclecoeff\right\|\leq \beta_3$, for some $\beta_1$, $\beta_2$, $\beta_3>0$.

Let a Lyapunov candidate be $\LyapunovTotal = (1-d)\LyapunovR + d\LyapunovV$ with some constant $d\in(0,1)$ to be chosen. Then along the trajectories of the coupled system, we have
\begin{equation*}
	\begin{aligned}
			\frac{\ud \LyapunovTotal}{\ud t} &= (1-d)\frac{\partial\LyapunovR}{\partial x}f(x,y+h(x)) + d\frac{\partial \LyapunovV}{\partial y}\frac{\ud y}{\ud t} \\
			&= (1-d)\frac{\partial\LyapunovR}{\partial x}f(x,h(x)) + d\frac{\partial \LyapunovV}{\partial y}\frac{\ud g(x,y+h(x))}{\ud \varepsilon} \\
			&\quad + (1-d)\frac{\partial\LyapunovR}{\partial x}\left(f(x,y+h(x)) - f(x,h(x))\right) + d\frac{\partial \LyapunovV}{\partial y}\left(-\frac{\partial h}{\partial x}\right)f(x,y+h(x)) \\
			&\leq -(1-d)\xi\psi_1^2(x) - d\left(\lambda^2\psi_2^2(y) + k\tau\psi_3^2(y) + k\tauAdapt\psi_4^2(y)\right) \\
			&\quad + (1-d)\left\|\frac{\partial\LyapunovR}{\partial \ptheta}\right\| \left\|\left(\musclecoeff(x)u\right)_s - \left(\musclecoeff(x)\uref\right)_s\right\| + d\left\|\varepsilon y\right\|\left\|\frac{\partial h}{\partial \theta}\right\|\left\|\frac{1}{\varrho\Isma}\ptheta\right\| \\
			&\leq -(1-d)\xi\psi_1^2(x) - d\left(\lambda^2\psi_2^2(y) + k\tau\psi_3^2(y) + k\tauAdapt\psi_4^2(y)\right) \\
			&\quad + (1-d)\psi_1(x) \left(\beta_2\psi_3(y)+\beta_3\psi_2(y)\right) + d\left(\tau\psi_3(y)+\tauAdapt\psi_4(y)\right)\beta_1\psi_1(x) \\
			&\leq -\psi^\top\Lambda\psi
		\end{aligned}
\end{equation*}
where $\psi=\left(\begin{smallmatrix}
		\psi_1 \\ \psi_2 \\ \psi_3 \\ \psi_4
	\end{smallmatrix}\right)$ and 
\begin{equation}
	\Lambda = \begin{bmatrix}
			(1-d)\xi 				& -\frac{(1-d)}{2}\beta_3 & -\frac{(1-d)}{2}\beta_2-\frac{d}{2}\tau\beta_1 & -\frac{d}{2}\tauAdapt\beta_1 \\
			-\frac{(1-d)}{2}\beta_3 & d\lambda^2 		      & 0 & 0 \\
			-\frac{(1-d)}{2}\beta_2-\frac{d}{2}\tau\beta_1 & 0 & dk\tau & 0 \\
			-\frac{d}{2}\tauAdapt\beta_1 & 0 & 0 & dk\tauAdapt
		\end{bmatrix}
\end{equation}
Note that the right-hand side of the inequality $\frac{\ud \LyapunovTotal}{\ud t}\leq -\psi^\top\Lambda\psi$ is quadratic in $\psi$ which is negative definite when matrix $\Lambda$ is positive definite. The determinant of $\Lambda$ is given by
\begin{equation}
	\begin{aligned}
			|\Lambda| &= (1-d)d^3\xi\lambda^2k^2\tau\tauAdapt - \left(\frac{(1-d)}{2}\beta_3\right)^2d^2k^2\tau\tauAdapt \\
			&\quad - \left(\frac{(1-d)}{2}\beta_2+\frac{d}{2}\tau\beta_1\right)^2d^2\lambda^2k\tauAdapt - \left(\frac{d}{2}\tauAdapt\beta_1\right)^2d^2\lambda^2k\tau \\
			&= -a_1k\left(\tau^2\tauAdapt + \tau\tauAdapt^2\right) -a_2k\tau\tauAdapt + a_3k^2\tau\tauAdapt - a_4k\tauAdapt
		\end{aligned}
\end{equation}
where $a_1=\frac{d^4}{4}\lambda^2\beta_1^2$, $a_2=\frac{(1-d)d^3}{2}\lambda^2\beta_1\beta_2$, $a_3=(1-d)d^3\xi\lambda^2-\frac{(1-d)^2d^2}{4}\beta_3^2$, and $a_4=\frac{(1-d)^2d^2}{4}\lambda^2\beta_2^2$. Note that we can write $\tauAdapt=q_1\tau$ with some factor $q_1>0$. Then, from Lemma~\ref{lemma:cable_stability}, the gain $k=\frac{q_2}{\tau}$ where $q_2=\frac{1+q_1-\sqrt{(1+q_1)^2-q_1(4-(1+b)^2)}}{2q_1}$. Then, we have
\begin{equation}
	|\Lambda| = -a_1(1+q_1)q_1q_2\tau^2 - a_2q_1q_2\tau + (a_3q_2-a_4)q_1q_2
\end{equation}
The matrix $\Lambda$ being positive definite is equivalent to
\begin{equation}
	\tau < \frac{-a_2+\sqrt{a_2^2+4a_1(1+q_1)(a_3q_2-a_4)}}{2a_1(1+q_1)} =: \tau_d
\end{equation}
where the right-hand side of the inequality is a function of the parameter $d$ in the Lyapunov candidate $\LyapunovTotal$. An additional constraint is that $a_3q_2-a_4> 0$ which is given by
\begin{equation}
	\begin{aligned}
			& a_3q_2-a_4 = q_2\left((1-d)d^3\xi\lambda^2-\frac{(1-d)^2d^2}{4}\beta_3^2\right) - \frac{(1-d)^2d^2}{4}\beta_2^2\lambda^2 > 0 \\
			\Rightarrow\quad & d > \frac{q_2\beta_3^2+\lambda^2\beta_2^2}{q_2\beta_3^2+\lambda^2\beta_2^2 + 4q_2\xi\lambda^2}
		\end{aligned}
\end{equation}

Thus we finally obtain $\frac{\ud \LyapunovTotal}{\ud t}\leq 0$. Furthermore, $\frac{\ud \LyapunovTotal}{\ud t} = 0$ only at the equilibrium of the coupled system. Hence, we proved that the coupled system is also (locally) asymptotically stable at the equilibrium.

\section{Weight function $\weight$ for neural ring model}\label{appdx:weight}

The weight function is given by
\begin{equation}
	\weight(\varphi) = \sum_{n=-5}^{5} \hat{\weight}_n e^{jn\varphi}
\end{equation}
The coefficients are given by $\hat{\weight}_n=\frac{\hat{\mathsf{V}}_n\hat{\mathsf{f}}_n}{0.01+|\hat{\mathsf{f}}_n|^2}$ where $\hat{\mathsf{f}}$ and $\hat{\mathsf{V}}_n$ are the fourier coefficients of the functions $\mathsf{f}^{\text{desired}}(\cdot)$ and $\mathsf{V}^{\text{desired}}(\cdot)$, respectively. The function $\mathsf{f}^{\text{desired}}(\cdot)$ is 
given by
\begin{equation}
	\mathsf{f}^{\text{desired}}(\varphi) = 2.53+34.8e^{8.08(\cos(\varphi)-1)}
\end{equation}
The function $\mathsf{V}^{\text{desired}}(\cdot)$ 
is given by
\begin{equation}
	\mathsf{V}^{\text{desired}}(\varphi) = h^{-1}(\mathsf{f}^{\text{desired}}(\varphi))
\end{equation}
where $h(\cdot)$ is given in~\eqref{eq:synaptic_response-numeric}. The details of the weight function derivation can be found in~\cite{zhang1996representation}.

\section{Explicit formulae of consensus algorithm}\label{appdx:sensing_algorithm}

\noindent
\textbf{Explicit formula of the update rule for \textit{proprioception}~\eqref{eq:theta_neural_ring_update}:}
\begin{equation}
	\gamma^{\uptheta_i} = \left\{\begin{aligned}
		\tauRing \frac{k_\uptheta}{2}\left(\sin(\hat{\uptheta}_i - \hat{\uptheta}_{i-1} - \upkappa_i\Delta s_{i,i-1}) - \sin(\hat{\uptheta}_{i+1} - \hat{\uptheta}_i - \upkappa_{i+1}\Delta s_{i+1,i}) \right),\quad& i=1,\ldots,N-1, \\
		\tauRing\frac{k_\uptheta}{2}\left(\sin(\hat{\uptheta}_{N} - \hat{\uptheta}_{N-1} - \upkappa_{N}\Delta s_{N,N-1})\right),\quad& i=N.
	\end{aligned}\right.
	\label{eq:consensus_dynamics_theta}
\end{equation}
In numerical simulation, the midpoint rule is used for curvature inputs, i.e., $\bar{\upkappa}_i=\frac{1}{2}(\upkappa_i+\upkappa_{i-1})$ instead of $\upkappa_i$, to increase the accuracy of the discrete approximation.

\noindent
\textbf{Explicit formulae of the update rule for \textit{chemosensing}~\eqref{eq:alpha_neural_ring_update}:} 
\begin{subequations}
	\begin{align}
		\gamma^{\upalpha_i}  &= \tauRing k_{\rsucker} e^{-\hat{\upmu}_i\c_i} \sum_{j\in \mathcal{N}_i} \left(\rt_i(\hat{\upalpha}_i,\hat{\upmu}_i,\hat{\uptheta}_i; \c_i) - \rt_j(\hat{\upalpha}_i,\hat{\upmu}_i,\hat{\uptheta}_i; \c_j) \right)\begin{bmatrix}
			-\sin(\hat{\upalpha}_i+\hat{\uptheta}_i) \\ \cos(\hat{\upalpha}_i+\hat{\uptheta}_i)
		\end{bmatrix} - \gamma^{\uptheta_i} 
		\label{eq:consensus_dynamics_alpha} \\
		\frac{\ud \hat{\upmu}_i}{\ud t} &= \c_ie^{-\hat{\upmu}_i\c_i}\sum_{j\in \mathcal{N}_i} k_{\rsucker}\left(\rt_i(\hat{\upalpha}_i,\hat{\upmu}_i,\hat{\uptheta}_i; \c_i) - \rt_j(\hat{\upalpha}_i,\hat{\upmu}_i,\hat{\uptheta}_i; \c_j) \right)\begin{bmatrix}
			\cos(\hat{\upalpha}_i+\hat{\uptheta}_i) \\ \sin(\hat{\upalpha}_i+\hat{\uptheta}_i)
		\end{bmatrix} 
		\label{eq:consensus_dynamics_mu} \\
		&\quad - \sum_{j\in \mathcal{N}_i} k_\upmu\left(\hat{\upmu}_i - \hat{\upmu}_j\right) \notag
	\end{align}
	\label{eq:consensus_dynamics_bearing}
\end{subequations}

\smallskip
\noindent
\textbf{Approximation for the difference term in sensor positions $\sum_{j\in\mathcal{N}_i}\rsucker_i-\rsucker_j$:}
Notice that the explicit formulae of the update rule for \textit{chemosensing} include the global positions $\rsucker_i$'s of all the sensors which come from the definition of the chemosensing energy $\mathcal{E}^{\text{chemo}}$~(\ref{eq:consensus_cost_chemosensing},\,\ref{eq:target_vector_estimate}). This is the information that sensors do not have access to. However, these gradients only involve the differences between position vectors of neighboring sensors $\rsucker_i - \rsucker_j, j \in \mathcal{N}_i, i = 1,..., N$. These differences are approximated by the second-order derivative of the sensor location with respect to $s$ as follows:
\begin{equation}
	\begin{aligned}
		\sum_{j\in\mathcal{N}_i}\rsucker_i-\rsucker_j &= \rsucker_i - \rsucker_{i-1} + \rsucker_i - \rsucker_{i+1} = - \left[(\rsucker_{i+1} - \rsucker_i) - (\rsucker_i - \rsucker_{i-1})\right] \\
		&\approx \rsucker_{ss}|_{s=s_i} \approx \left\{\begin{aligned}
			-\begin{bmatrix}
				\cos(\hat{\uptheta}_i) \\ \sin(\hat{\uptheta}_i)
			\end{bmatrix}\Delta s, &\quad i=1, \\
			\begin{bmatrix}
				\sin(\hat{\uptheta}_i) \\ -\cos(\hat{\uptheta}_i)
			\end{bmatrix}\upkappa_i \Delta s^2, &\quad i=2,\ldots,N-1, \\
			\begin{bmatrix}
				\cos(\hat{\uptheta}_i) \\ \sin(\hat{\uptheta}_i)
			\end{bmatrix}\Delta s, &\quad i=N.
		\end{aligned}\right.
	\end{aligned}
\end{equation}
where the differences are approximated only with local information including the shape angle estimates $\hat{\uptheta}_i$ and the curvature inputs $\upkappa_i$.

\section{Proof of Theorem~\ref{thm:consensus}} \label{appdx:consensus_proof}

According to~\eqref{eq:angular_velocity-shifting-mapping}, we have
\begin{equation}
	\frac{\ud \hat{\uptheta}_i(t)}{\ud t} = -\gamma^{\uptheta_i}(t)/\tauRing,\quad  \frac{\ud \hat{\upalpha}_i(t)}{\ud t} = -\gamma^{\upalpha_i}(t)/\tauRing,\quad \text{for all }t\geq0
\end{equation}

\subsection{Proof of Part A (Proprioception)}\label{appdx:proprioception_proof}
From the explicit formula of the update rule~\eqref{eq:consensus_dynamics_theta} for proprioception, the solution $\bm{\hat{\uptheta}^\star}:=(\hat{\uptheta}_1^\star, \ldots, \hat{\uptheta}_N^\star)$ is an equilibrium. We calculate the Jacobian matrix at that equilibrium:
\begin{equation}
	\mathsf{A} := \frac{\partial}{\partial \bm{\hat{\uptheta}}}\left(\frac{\ud \bm{\hat{\uptheta}}}{\ud t}\right)\Big|_{\bm{\hat{\uptheta}} = \bm{\hat{\uptheta}^\star}} = \frac{k_\uptheta}{2}\begin{bmatrix}
			-2 &  1 &    &        &    &    &    \\
			1  & -2 &  1 &        &    &    &    \\
			   &    &    & \ddots &    &    &    \\
			   &    &    &        &  1 & -2 &  1 \\
			   &    &    &        &    &  1 & -1 \\
		\end{bmatrix}
\end{equation}
Since $\mathsf{A}$ is symmetric and negative definite, all its eigenvalues are real and negative. Then, the equilibrium $\bm{\hat{\uptheta}}=\bm{\hat{\uptheta}^\star}$ is (locally) asymptotically stable by~\cite[Chap. 4-Theorem 4.7]{khalil2002nonlinear}. In numerical simulation, we set $\hat{\uptheta}_1\equiv \hat{\uptheta}_1^\star = 0$, following the hard constraint of fixed orientation of the arm at the base.

\begin{remark}
	Furthermore, note that for every $\bm{\hat{\uptheta}}$ where its $i$-th entry $\hat{\uptheta}_i=\hat{\uptheta}_i^\star + \mathsf{n}_i\pi$ for $\mathsf{n}_i\in\{-1,0,1\}$, it is also an equilibrium of the dynamics. However, for each equilibrium that has at least one entry where $\mathsf{n}_i\neq0$, the Jacobian matrix at that equilibrium has one or more positive eigenvalues, which indicates that such equilibrim is unstable (the detailed proof of which is omitted here). Thus, all equilibria are unstable except the desired equilibrium $\bm{\hat{\uptheta}}=\bm{\hat{\uptheta}^\star}$ which corresponds to $\mathsf{n}_i=0$ for all $i$.
\end{remark}

\subsection{Proof of Part B (Chemosensing)} \label{appdx:chemosensing_proof}

\begin{lemma}
	Consider the dynamics~\eqref{eq:consensus_dynamics_alpha} with constants $\upmu_i=\mu$, $\c_i=\c_i^\star$, and $\hat{\uptheta}_i=\uptheta_i$, i.e., the only state is $\hat{\upalpha}_i$, for $i=1,\ldots,N$. Then the dynamics is (locally) asymptotically stable at the equilibrium $\bm{\hat{\upalpha}}=\bm{\upalpha}^\star+\bm{\uptheta}^\star - \bm{\hat{\uptheta}}^\star=:\bm{\hat{\upalpha}}^\star$.
	\label{lemma:alpha_stability}
\end{lemma}

\begin{proof}
	With the constraints of $\upmu_i=\mu$, $\c_i=\c_i^\star$, and $\hat{\uptheta}_i=\uptheta_i$, the energy for chemosensing~\eqref{eq:consensus_cost_chemosensing} is a function of only bearing estimates $\bm{\hat{\upalpha}}$, i.e., $\mathcal{E}^{\text{chemo}}=\mathcal{E}^{\text{chemo}}(\bm{\hat{\upalpha}})$. It is positive and only takes zero when $\rt_i=\rt_j=:\rt$ for all $i,j=1,\ldots,N$. Since $\upmu_i=\mu$, $\c_i=\c_i^\star$, we have $\hat{\uprho}_i=\uprho_i$ which implies $|\rt-\rsucker|=|\r^\target-\rsucker|$ for $i=1,\ldots,N$. Since the equation holds for all $\rsucker$'s, we have a unique solution to the target estimate $\rt=\r^\target$ which corresponds to $\bm{\hat{\uptheta}}^\star+\bm{\hat{\upalpha}}=\bm{\uptheta}^\star+\bm{\upalpha}^\star$. Thus, the function $\mathcal{E}^{\text{chemo}}$ is positive and only takes zero at the equilibrium $\bm{\hat{\upalpha}}=\bm{\hat{\upalpha}}^\star$, hence a Lyapunov candidate function. Next, we investigate the time derivative of that function.
	\begin{equation}
		\begin{aligned}
			\frac{\ud}{\ud t}\mathcal{E}^{\text{chemo}} (\bm{\hat{\upalpha}}) &= \sum_{i=1}^{N} \nabla_{\hat{\upalpha}_i}\mathcal{E}^{\text{chemo}}\frac{\ud \hat{\upalpha}_i}{\ud t} = -\sum_{i=1}^N \left(\nabla_{\hat{\upalpha}_i}\mathcal{E}^{\text{chemo}}\right)^2
		\end{aligned}
	\end{equation}
	The explicit formula of the gradient $\nabla_{\hat{\upalpha}_i}\mathcal{E}^{\text{chemo}}$ takes zero only at $\rt_i=\rt_j=\rt$ and thus at the equilibrium $\bm{\hat{\upalpha}}=\bm{\hat{\upalpha}}^\star$ as described above. Thus, the time derivative of the energy function is negative and only takes zero at the equilibrium. By Lyapunov theory, the equilibrium is (locally) asymptotically stable.
\end{proof}

\smallskip
By taking $k_\uptheta$ to be large enough, we can regard the proprioception dynamics of $\bm{\hat{\uptheta}}$ as `fast' system. The cheomsensing dynamics of $\bm{\hat{\upalpha}}$ is considered as `slow' system. The fast system is asymptotically stable at the equilibrium $\bm{\hat{\uptheta}}=\bm{\hat{\uptheta}}^\star$ as proved in Appendix~\ref{appdx:proprioception_proof}. The slow system in the manifold $\{(\bm{\hat{\uptheta}}, \bm{\hat{\upalpha}}): \bm{\hat{\uptheta}}=\bm{\hat{\uptheta}}^\star\}$ is also proved to be (locally) asymptotically stable at the equilibrium $\bm{\hat{\upalpha}}=\bm{\hat{\upalpha}}^\star$ by Lemma~\ref{lemma:alpha_stability}. Then, similar proof can be derived using the method of singular pertuerbation as in Appendix~\ref{appdx:sensoryfeedback_control_dynamics_proof} for Theorem~\ref{thm:sensoryfeedback_control} part B. Further details of the proof is omitted.

\begin{figure*}[t]
	\centering
	\includegraphics[width=\textwidth, trim = {10pt 10pt 10pt 10pt}, clip = true]{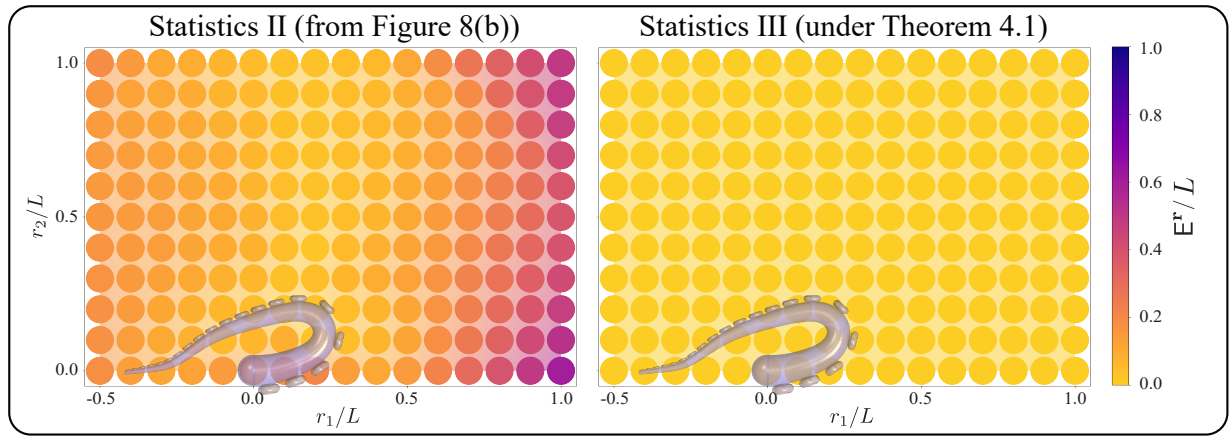}
	\vspace{-15pt}
	\caption{Sensing statistics of a bent arm. On the left is the Statistics II from Figure~\ref{fig:sensing_simulation}.(b) where sensory inputs and $\hat{\upmu}_i(t)$ are dynamic. On the right is the Statistics III under all the assumptions in Theorem~\ref{thm:consensus} where sensory inputs are steady and $\hat{\upmu}_i=\mu$ for $i=1,\ldots,N$.}
	\label{fig:sensing_statistics}
	\vspace{0pt}
\end{figure*}

\section{Statistics of bent arm sensing under Theorem~\ref{thm:consensus}}\label{appdx:statistics}

Theorem~\ref{thm:consensus} is further verified in numerical simulation through the performance of the sensing algorithm. Similar to the setup of Statistics II discussed in \S\,\ref{sec:sensing_results}, a total of 176 targets are selected on a uniform grid inside a rectangle $[-0.5L,L]\times[0,L]$ for a bent arm. The error statistics (named Statistics III) is illustrated in Figure~\ref{fig:sensing_statistics} on the right with Statistics II shown on the left for comparison.
The min, max and mean values of $\Er/L$ are $1.27\times10^{-4}$, $7.70\times10^{-3}$ and $2.19\times10^{-3}$, respectively. 

It is observed that the sensing algorithm achieves high level of accuracy under the assumptions in Theorem~\ref{thm:consensus}. This is because static $\c_i$ and $\hat{\upmu}_i$ eliminate the error in range estimation $\hat{\uprho}_i$ which avoids the problem of dilution of precision (See Remark~\ref{remark:mu_estimate}). Thus, the algorithm is able to achieve low error as shown in Statistics III, which further supports the asymptotic stability of the equilibrium in Theorem~\ref{thm:consensus}.

\bibliographystyle{ieeetr}
\bibliography{bibfiles/octopus_papers,bibfiles/reference,bibfiles/motivation}

\end{document}

%% file: _macros.tex
\usepackage{amsfonts, dsfont, mathtools, amsmath, amssymb, bbold, mathrsfs, bm, upgreek}
\usepackage{graphicx}     
\usepackage[table]{xcolor} 
\usepackage{times, tabularx, subfigure, float} 
\usepackage{nicefrac, multirow, multicol, arydshln, multimedia}

\usepackage{algorithm}
\usepackage{algpseudocode}

\graphicspath{{figures/}}

\usepackage{titlesec}
\renewcommand\thesubsection{\arabic{section}.\arabic{subsection}}

\usepackage{amsthm}
\usepackage{setspace}
\usepackage[left=1in, right=1in, top=1in, bottom=1in]{geometry}
\usepackage[font=small]{caption}

%% file: _definitions.tex
\def\R{{\mathds{R}}}

\def\0{{\mathbb{0}}}\label{key}
\def\1{{\mathds{1}}}

\newcommand{\abs}[1]{\left| #1 \right|}

\definecolor{db}{RGB}{23,20,119}
\definecolor{dg}{RGB}{2,101,15}

\newtheorem{theorem}{Theorem}[section]

\newtheorem{lemma}{Lemma}[section]

\newtheorem{definition}{Definition}
\newtheorem{remark}{Remark}

\newcommand{\dif}{\mathrm{d}}

\newcommand{\set}[1]{\left\{#1\right\}}

\newcommand{\material}[1]{
	\ifthenelse{\equal{#1}{\kappa}}{\upkappa}{
		\ifthenelse{\equal{#1}{\nu}}{\upnu}{
			\ifthenelse{\equal{#1}{\omega}}{\upomega}{
				\ifthenelse{\equal{#1}{\sigma}}{\upsigma}{
					\ifthenelse{\equal{#1}{\theta}}{\uptheta}{
						\mathsf{#1}}}}}}
}

\newcommand{\muscle}{\text{m}}
\newcommand{\longitudinalmuscle}{\text{LM}}
\newcommand{\transversemuscle}{\text{TM}}

\newcommand{\ud}{\,\mathrm{d}}

\DeclareMathOperator*{\argmin}{arg\,min}
\DeclareMathOperator*{\argmax}{arg\,max}
\DeclareMathOperator\arctanh{arctanh}

\newcommand{\myper}{_\text{per}}
\newcommand{\mytan}{_\text{tan}}
\newcommand{\rhow}{\varrho_\text{water}}
\newcommand{\target}{\text{target}}

\renewcommand{\c}{\mathsf{c}}
\renewcommand{\r}{\mathbf{r}}
\newcommand{\rt}{\hat{\bm{\mathsf{r}}}^{\text{target}}}
\newcommand{\rsucker}{\bm{\mathsf{r}}}

\newcommand{\radius}{r^\text{rod}}
\newcommand{\rodtip}{r^\text{tip}}
\newcommand{\rodbase}{r^\text{base}}

\newcommand{\dist}{\rho}

\newcommand{\unicycle}[1]{\mathsf{#1}}
\newcommand{\thetaUnicycle}{\vartheta}

\newcommand{\Hamiltonian}{\mathcal{H}}
\newcommand{\potential}{\mathcal{V}}
\newcommand{\kinetic}{\mathcal{T}}

\newcommand{\storedenergy}{\mathcal{W}}

\newcommand{\Vadapt}{W}
\newcommand{\tauAdapt}{\tilde{\tau}}
\newcommand{\tauRing}{\uptau}

\newcommand{\LM}{\longitudinalmuscle}
\newcommand{\LMt}{{\longitudinalmuscle_\mathsf{t}}}
\newcommand{\LMb}{{\longitudinalmuscle_\mathsf{b}}}
\newcommand{\TM}{\transversemuscle}

\newcommand{\Vtop}{V^{\LMt}}
\newcommand{\Vbot}{V^{\LMb}}
\newcommand{\Vmuscle}{V^{\muscle}}

\newcommand{\Imuscle}{I^{\muscle}}

\newcommand{\VmA}{W^{\muscle}}
\newcommand{\lambdaA}{\hat{\lambda}}
\newcommand{\Lyapunov}{\mathcal{E}}

\newcommand{\LyapunovR}{\Hamiltonian}
\newcommand{\LyapunovV}{\Lyapunov_0}
\newcommand{\LyapunovTotal}{\Lyapunov}

\newcommand{\Vref}{\tilde{V}}
\newcommand{\Wref}{\tilde{W}}
\newcommand{\Nelem}{N^{\text{arm}}}
\newcommand{\Ncable}{N^{\text{cable}}}
\newcommand{\Nring}{N^{\text{ring}}}
\newcommand{\weight}{\mathsf{W}}
\newcommand{\damping}{\xi}
\newcommand{\musclecoeff}{\mathsf{v}}
\newcommand{\const}{\varsigma}
\newcommand{\Isma}{\mathrm{I}}
\newcommand{\Er}{\mathsf{E}^{\r}}
\newcommand{\uref}{\tilde{u}}
\newcommand{\pr}{\mathbf{p}}
\newcommand{\ptheta}{\mathsf{p}}